\documentclass[final,3p,times]{article}

\usepackage[a4paper,left=22mm,top=20mm,right=22mm,bottom=35mm]{geometry}

\usepackage{authblk}
\usepackage[numbers, sort&compress]{natbib}
\usepackage[margin=30pt,font=small,labelfont=bf]{caption}
\usepackage{amsmath,amsopn,amsthm,amssymb}

\usepackage[colorlinks]{hyperref}
\hypersetup{
		citecolor=blue,
    linkcolor=blue,
}
\usepackage{graphicx}
\usepackage{mathptmx} 
\usepackage[mathscr]{euscript}
\usepackage{natbib}
\usepackage{mathtools}
\usepackage{amssymb}
\usepackage{wasysym}
\usepackage{float}
\usepackage{enumitem}
\usepackage{xspace}
\usepackage{url}

\usepackage{algorithm}
\usepackage[noend]{algpseudocode}
\algrenewcommand\algorithmicrequire{\textbf{Input:}}
\algrenewcommand\algorithmicensure{\textbf{Output:}}

\newtheorem{theorem}{Theorem}[section]
\newtheorem{lemma}[theorem]{Lemma} 

\newtheorem{proposition}[theorem]{Proposition}
\newtheorem{example}[theorem]{Example}
\newtheorem{definition}[theorem]{Definition}
\newtheorem{observation}[theorem]{Observation}

\newtheorem{remark}[theorem]{Remark}

\usepackage{color}
\usepackage[dvipsnames]{xcolor}

\definecolor{orange}{RGB}{255, 171, 0}
\definecolor{petrol}{RGB}{0, 95, 105}
\makeatletter
\def\moverlay{\mathpalette\mov@rlay}
\def\mov@rlay#1#2{\leavevmode\vtop{%
   \baselineskip\z@skip \lineskiplimit-\maxdimen
   \ialign{\hfil$\m@th#1##$\hfil\cr#2\crcr}}}
\newcommand{\charfusion}[3][\mathord]{
    #1{\ifx#1\mathop\vphantom{#2}\fi
        \mathpalette\mov@rlay{#2\cr#3}
      }
    \ifx#1\mathop\expandafter\displaylimits\fi}
\makeatother

\newcommand{\cupdot}{\charfusion[\mathbin]{\cup}{\cdot}}

\makeatletter
\newcommand{\cuptimes}{\mathbin{\cup\mkern-11.6mu\mathpalette\@cuptimes\relax}}
\newcommand{\@cuptimes}[2]{%
  \sbox0{$\m@th#1\cup$}%
  \sbox2{$\m@th#1\scriptstyle\times$}%
  \raisebox{0.2\dimexpr\ht0-\ht2}{\box2}%
}
\makeatother

\newcommand{\join}{\cuptimes} % Adjusted \cuptimes command

\DeclareMathOperator*{\argmax}{arg\,max}

\DeclareMathOperator{\child}{child}
\newcommand{\lca}{\ensuremath{\operatorname{lca}}}

\newcommand{\parent}{\ensuremath{\operatorname{par}}}
\newcommand{\gatex}{\textsc{GaTEx}\xspace}

\newcommand{\MD}{\ensuremath{\mathbb{M}}}
\newcommand{\MDstrong}{\ensuremath{\mathbb{M}_{\mathrm{str}}}}
\newcommand{\Mmax}{\ensuremath{\mathbb{M}_{\max}}}
\newcommand{\MDT}{\ensuremath{\mathscr{T}}}

\usepackage{algorithm}
\usepackage[noend]{algpseudocode}
\usepackage{algpseudocodex}
\algrenewcommand\algorithmicrequire{\textbf{Input:}}
\algrenewcommand\algorithmicensure{\textbf{Output:}}
\usepackage{eqparbox,array}

\newcommand{\omegaExclHyb}{\ensuremath{\omega_{\neg\eta}}}

\providecommand{\keywords}[1]{\textbf{\textit{Keywords: }} #1}

\title{Solving NP-hard Problems on \textsc{GaTEx} Graphs: Linear-Time Algorithms for Perfect Orderings, Cliques, Colorings, and Independent Sets\thanks{This contribution is an
    extended version of the COCOON'23 paper \cite{HS-GatexLinTime:23}.}}

\author[1]{Marc Hellmuth} 
\author[2]{Guillaume E. Scholz} 

\affil[1]{Department of Mathematics, Faculty of Science,
  Stockholm University, SE-10691 Stockholm, Sweden 
  \newline \texttt{marc.hellmuth@math.su.se}}

\affil[2]{Bioinformatics Group, Department of Computer Science \&
    Interdisciplinary Center for Bioinformatics, Universit{\"a}t Leipzig,
    H{\"a}rtelstra{\ss}e~16--18, D-04107 Leipzig, Germany.}

\date{\ }

\begin{document}
\sloppy

\maketitle

\abstract{
The class of \textsc{Ga}lled-\textsc{T}ree \textsc{Ex}plainable (\textsc{GaTEx}) graphs has recently been discovered as a natural generalization of cographs. Cographs are precisely those graphs that can be uniquely represented by a rooted tree where the leaves correspond to the vertices of the graph. As a generalization, \textsc{GaTEx} graphs are precisely those that can be uniquely represented by a  particular rooted acyclic network, called a galled-tree.

This paper explores the use of galled-trees to solve combinatorial problems on \textsc{GaTEx} graphs that are, in general, NP-hard. We demonstrate that finding a maximum clique, an optimal vertex coloring, a perfect order, as well as a maximum independent set in \textsc{GaTEx} graphs can be efficiently done in linear time. The key idea behind the linear-time algorithms is to utilize the galled-trees that explain the \textsc{GaTEx} graphs as a guide for computing the respective cliques, colorings, perfect orders, or independent sets. }

\smallskip
\noindent

\section{Introduction}

Modular decomposition is a general technique to display nested ``substructures'' (modules) of a given
graph in the form of a rooted tree (the \emph{modular decomposition tree of $G$}) whose inner
vertices are labeled with ``0'', ``1'', and ``prime''. Cographs are precisely those graphs for which the
modular decomposition tree has no prime vertices. In this case, complete structural information of
the underlying cograph, i.e., the knowledge of whether two vertices are linked by an edge or not, is
provided by the modular decomposition tree. As a consequence, these modular decomposition trees
serve as a perfect guide for algorithms to efficiently solve many computationally hard problems on
cographs (e.g., the graph-isomorphism problem or classical NP-hard problems like ``minimum
independent set'', ``maximum clique'', or ``minimum vertex coloring'') \cite{Corneil:81,Corneil:85}. 
However, when encountering prime vertices, conventional modular decomposition trees do
not provide full structural information about the underlying graphs and become less useful for 
algorithmic solutions to hard problems.  To circumvent this issue, we aim at using modular
decomposition \emph{networks} instead of trees. In \cite{HS:22}, we focused on particular
networks, called  galled-trees, that are obtained from the modular decomposition tree by replacing prime vertices by
rooted 0/1-labeled cycles. A graph $G = (X,E)$ is \emph{{\textsc{Ga}}lled-{\textsc{T}}ree
{\textsc{Ex}}plainable} (\gatex) if there is a 0/1-labeled galled-tree $(N,t)$ such that ${x,y}\in
E$ if and only if the label $t(\lca_N(x,y))$ of the unique least-common ancestor of $x$ and $y$ in
$N$ is ``1''. \gatex graphs, thus, naturally generalize the concept of cographs.
Further exploration of the class of \gatex graphs in \cite{HS-GatexForbSubg:23} shows that these
graphs are characterized by the absence of 25 forbidden subgraphs. This, in turn, implies that
\gatex graphs are closely linked to other famous graph classes such as weakly-chordal graphs,
perfect graphs with perfect order, comparability and permutation graphs, murky graphs as well as
interval graphs, Meyniel graphs, or very strongly-perfect and brittle graphs. In addition, every
\gatex graph has twin-width at most 1.

Cotrees serve as a guide for algorithms on cographs to solve many combinatorial problems that are
classified as ``hard''. In this contribution, we ask whether the galled-trees that explain \gatex
graphs can be used in a similar manner. In particular, we are interested in the following classical
NP-hard problems \cite{garey1979computers}: Determining the size $\omega(G)$ of a maximum clique
and finding such a clique, the size $\chi(G)$ of an optimal vertex-coloring and finding such a
coloring, the size $\alpha(G)$ of a maximum independent set of a given graph $G$ and finding such an
independent set. In general, determining the invariants $\omega(G)$, $\chi(G)$, and $\alpha(G)$ for
arbitrary graphs $G$, as well as finding the underlying optimal cliques, colorings, and independent
sets, is an NP-hard task \cite{garey1979computers}. All these invariants are not only of interest
from a theoretical point of view but also have many practical applications in case the underlying
graph models real-world structures, e.g., social networks \cite{Luce1949}, gene/protein-interaction
networks \cite{BSY:99,VL:03}, job/time-slots assignments in scheduling problems
\cite{Marx2004GRAPHCP}, and many more. In addition, we consider the problem of determining a perfect
ordering of \gatex graphs, i.e., an ordering of the vertices of $G$ such that a greedy coloring
algorithm with that ordering optimally colors every induced subgraph of $G$. As shown by Middendorf
and Pfeiffer \cite{MIDDENDORF1990327}, the problem of deciding whether a graph is perfectly
orderable is NP-complete. As we will argue below, the problem of finding a perfect ordering remains
NP-hard even for perfectly orderable graphs.

We show here that $\omega(G)$, $\chi(G)$, $\alpha(G)$ as well as a perfect ordering can be computed
in linear time for \gatex graphs $G$. The crucial idea for the linear-time algorithms is to avoid
working directly on the \gatex graphs $G$ but rather to utilize the galled-trees that explain $G$ as
a guide for the algorithms to compute these invariants. In particular, we show first how to employ
the galled-tree structure to compute a perfect ordering of \gatex graphs. This result is then used
to determine $\omega(G)$, $\chi(G)$, $\alpha(G)$. In addition, we provide algorithms
to find a maximum clique, an optimal vertex coloring as well as a maximum independent set
in \gatex graphs in linear-time. 

\section{Preliminaries}

\paragraph{\bf Graphs.}
We consider graphs $G=(V,E)$ with vertex set $V(G)\coloneqq V\neq \emptyset$ and
edge set $E(G)\coloneqq E$. A graph $G$ is \emph{undirected} if $E$ is a subset
of the set of two-element subsets of $V$ and $G$ is \emph{directed} if
$E\subseteq V\times V \setminus \{(v,v)\mid v\in V\}$. Thus, edges $e\in E$
in an undirected graph $G$ are of the form $e=\{x,y\}$ and in directed graphs of
the form $e=(x,y)$ with $x,y\in V$ being distinct. We write $H \subseteq G$ if
$H$ is a subgraph of $G$ and $G[W]$ for the subgraph in $G$ that is induced by
some subset $W \subseteq V$. A $P_4$ denotes an induced undirected path on four vertices.
We often write $a-b-c-d$ for an induced		 $P_4$ with vertices $a,b,c,d$ and edges 
$\{a,b\}, \{b,c\},\{c,d\}$.
An undirected graph is \emph{connected} if, for
every two vertices $u,v\in V$, there is a path connecting $u$ and $v$. A
directed graph is \emph{connected} if its underlying undirected graph is
connected. A (directed or undirected) graph $G$ is \emph{biconnected} if it
contains no vertex whose removal disconnects $G$. A \emph{biconnected component}
of a $G$ is an inclusion-maximal biconnected subgraph. If such a biconnected component is
not a single vertex or an edge, then it is called \emph{non-trivial}.
\smallskip

\begin{remark}
\emph{From here on, we will call an undirected graph simply \emph{graph}.}
\end{remark}

For two graphs $G$ and $H$ we put $G-H\coloneqq (V(G)\setminus V(H),
E(G)\setminus F)$ with $F\subseteq E(G)$ being the collection of all edges incident
to vertices in $V(H)$, and $G\cap H\coloneqq (V(G)\cap V(H), E(G)\cap E(H))$. For two vertex-disjoint
graphs $G$ and $H$, their \emph{disjoint union} is defined as $G\cupdot H \coloneqq (V(G)\cupdot
V(H), E(G)\cupdot E(H))$ while their \emph{join union} is defined as $G\join H \coloneqq
(V(G)\cupdot V(H), E(G)\cupdot E(H)\cupdot \{\{x,y\}\mid x\in V(G), y\in V(H)\})$.

A \emph{clique} of a graph $G$ is an inclusion-maximal complete subgraph $G$.
The size of a maximum clique of $G$ is called the \emph{clique number} and
denoted by $\omega(G)$. 
A \emph{coloring} of a graph $G$ is a map $\sigma\colon
V(G)\to S$, where $S$ denotes a set of colors, such that $\sigma(u)\neq
\sigma(v)$ for all $\{u,v\}\in E(G)$. The minimum number of colors needed for a
coloring of $G$ is called the \emph{chromatic number} of $G$ and denoted by
$\chi(G)$.
  A subset $W\subseteq V(G)$ of pairwise non-adjacent vertices is
called \emph{independent set}. The size of a maximum independent set in $G$ is
called the \emph{independence number} of $G$ and denoted by $\alpha(G)$. 
In general, determining the invariants $\omega(G)$, $\chi(G)$ and $\alpha(G)$ for arbitrary graphs is an
NP-hard task \cite{garey1979computers}.

A graph $G$ is \emph{perfect}, if the
chromatic number of every induced subgraph equals the size of the largest clique
of that subgraph. 
We consider total orders $\zeta = v_1\dots v_{|V|}$ defined on the vertex set $V$ of graphs $G=(V,E)$
and assume that  $v_i<v_j$ precisely if $v_i$ is left of $v_j$
in this sequence $\zeta$  (or equivalently, if $i<j$ in case indices are provided). We denote with $\zeta_{|H}$
the order $\zeta$ that is restricted to $V(H)$. 
Let $X$ and $Y$ be two disjoint subsets of $V(G)$. If $\zeta_1=x_1,x_2, \ldots
x_l$ and $\zeta_2=y_1,y_2, \ldots y_m$ are two total orderings on $X$ and $Y$,
respectively, then we denote with $\zeta_1\zeta_2$ the total ordering on $X \cup
Y$ given by concatenating $\zeta_1$ and $\zeta_2$, i.e.,
$\zeta_1\zeta_2=x_1,x_2, \ldots x_ly_1,y_2, \ldots y_m$. 

For a given total order $\zeta$ of $G$, a \emph{greedy coloring algorithm} scans the 
vertices in order $\zeta$ and assigns to each vertex $v$ the smallest positive
integer (color) assigned to none of the vertices $w<v$ that are adjacent to $v$.
A coloring of $G$ obtained with such an algorithm is called \emph{greedy coloring}.
A total order $\zeta$ of $G$ is \emph{perfect} if, for all induced subgraphs $H$ of $G$,
a greedy coloring algorithm that scans the vertices in order $\zeta_{|H}$ uses the minimum number of colors to color  $H$.
A graph $G$ is \emph{perfectly
orderable} if it admits a \emph{perfect order} $\zeta$. A total order $\zeta$ on $G$ contains an \emph{obstruction} (w.r.t. $G$) if there is an induced $P_4$
$a-b-c-d$ in $G$ such that $a<b$ and $c>d$ w.r.t.\ this order $\zeta$.
Every perfectly orderable graph is a perfect graph \cite{CHVATAL198463}.

\begin{proposition}[{\cite{CHVATAL198463}}]\label{prop:perfect-order}
A total order $\zeta$ on a graph $G$ is a perfect order if and only
if $\zeta$ does not contain any obstructions.
\end{proposition}

Perfectly orderable graphs are NP-complete to recognize \cite{MIDDENDORF1990327}. 
By Prop.\ \ref{prop:perfect-order}, one can test in polynomial-time as whether a given order is perfect:
simply check as whether one of the $O(|V|^4)$ induced $P_4$s yields an obstruction. 
This, in particular, implies that the problem to find a perfect ordering of a graph
remains NP-hard, even if the graph is already known to be perfectly orderable.

\paragraph{\bf Trees, Galled-trees and \gatex graphs.}
(Phylogenetic) trees and galled-trees are particular directed acyclic graphs
(DAGs). To be more precise, a {\em galled-tree} $N=(V,E)$ on $X$ is a DAG such that
either 
\begin{description}
	\item[\it (N0)] $V=X = \{x\}$ and, thus, $E=\emptyset$.
\end{description}
or $N$ satisfies  the following four properties 
\begin{description}%[noitemsep]
	\item[\it (N1)] There is a unique \emph{root} $\rho_N$ with indegree 0 and outdegree at least 2; and\smallskip
	\item[\it(N2)] $x\in X$ if and only if $x$ has outdegree 0 and indegree 1; and \smallskip
	\item[\it(N3)] Every vertex $v\in V^0 \coloneqq V \setminus X$ with $v\neq \rho_N$ has 
		\begin{enumerate}
			\item[(i)]indegree 1 and 	outdegree at least 2 (\emph{tree-vertex}) or
			\item[(ii)] indegree 2 and 	outdegree at least 1 (\emph{hybrid-vertex}).
		\end{enumerate}	
	\item[\it(N4)] Each biconnected component $C$ contains at most one hybrid-vertex $v$ 
					for which the two vertices $v_1,v_2$ with $(v_1,v), (v_2,v)\in E$
					belong to $C$. 
\end{description}
We note that in \cite{HS:22} galled-trees have been called level-1 networks. By
definition, every non-trivial biconnected component in a galled-tree $N$ forms
an (rooted) ``\emph{cycle}'' $C$ in $N$ \cite{HS18,CRV07} that is composed of
two directed paths $P^1(C)$ and $P^2(C)$ in $N$ (called \emph{sides} of $C$)
with the same start-vertex $\rho_C$ (the root of $C$) and end-vertex $\eta_C$
(the hybrid-vertex of $C$) and whose internal vertices, i.e., vertices in $C$ that are distinct from $\rho_C$ and $\eta_C$,
are pairwise distinct. 
\emph{Trees} are galled-trees without hybrid-vertices.
The leaf set $L(N)$ of a galled-tree $N$ is $X$, i.e., the set of all vertices satisfying \textit{(N2)}, 

Let $N=(V,E)$ be a galled-tree on $X$. A vertex $u\in V$ is called an
\emph{ancestor} of $v\in V$ and $v$ a \emph{descendant} of $u$, in symbols $v
\preceq_N u$, if there is a directed path (possibly reduced to a single vertex)
in $N$ from $u$ to $v$. We write $v \prec_N u$ if $v \preceq_N u$ and $u\neq v$.
If $(u,v)\in E$, then the vertex $v$ is a \emph{child of $u$} and $u$ is a
\emph{parent of $v$}. The set of children, resp., parents of a vertex $w$ in $N$
is denoted by $\child_N(w)$, resp., $\parent_N(w)$. For a non-empty subset of
leaves $A\subseteq X$, we define $\lca_N(A)$, or a \emph{lowest common
ancestor of $A$}, to be a $\preceq_N$-minimal vertex of $N$ that is an ancestor
of every vertex in $A$. For simplicity we put $\lca_N(x,y)\coloneqq
\lca_N(\{x,y\})$. 
By Lemma 49 and 67 in \cite{HSS:22cluster}, galled-trees $N$ are ``lca-networks'',
 i.e., $\lca_N(A)$ is uniquely determined for all $A \subseteq L(N)$.

We define $N(w)$ as the subgraph of $N$ rooted at $w$, i.e., the DAG
induced by $w$ and all its descendants. Morever, if the context is clear, 
we often write $L_w = L(N(w))$ for $w\in
V(N)$.

A galled-tree $N$ on $X$ is \emph{elementary} if it contains a single
rooted cycle $C$ of length $|X|+1$ with root $\rho_C = \rho_N$ and single
hybrid-vertex $\eta_C\in V(C)$ and additional edges $\{v_i,x_i\}$ such that
every vertex $v_i\in V(C)\setminus \{\rho_C\}$ is adjacent to a unique vertex
$x_i\in X	$. A galled-tree is \emph{strong} if it \emph{does not} contain cycles
of the following form: (i) $P^1(C)$ or $P^2(C)$ consist of $\rho_C$ and $\eta_C$
only or (ii) both $P^1(C)$ and $P^2(C)$ contain only one vertex distinct from
$\rho_C$ and $\eta_C$.

The tuple $(N,t)$ denotes a galled-tree $N=(V,E)$ on $X$ that is equipped with a
\emph{(vertex-)labeling} $t$ i.e., a map $t\colon V\to\{0,1,\odot\}$ such that
$t(x)=\odot$ if and only if $x\in X$. The graph $\mathscr{G}(N,t) = (X,E)$ with
vertex set $X$ and edges $\{x,y\}\in E$ precisely if $t(\lca_N(x,y))=1$ is said
to be \emph{explained} by $(N,t)$. A graph $G = (X,E)$ is
\emph{{\textsc{Ga}}lled-{\textsc{T}}ree {\textsc{Ex}}plainable (\gatex)}) if
there is a labeled galled-tree $(N,t)$ such that $G\simeq \mathscr{G}(N,t)$. A
labeling $t$ (or equivalently $(N,t)$) is \emph{quasi-discriminating} if
$t(u)\neq t(v)$ for all $(u,v)\in E$ with $v$ not being a hybrid-vertex. 
We note in passing, that quasi-discriminating labelings form a natural
generalization of discriminating labelings $t$ that require $t(u)\neq t(v)$
for all  $(u,v)\in E$ \cite{BD98}.

\begin{proposition}[{\cite{HS:22}}]
\gatex graphs can be recognized in linear-time and a galled-tree
$(N,t)$ that explains a \gatex graphs can be constructed in linear-time as well.
\end{proposition}
Moreover, \gatex graphs are characterized by a finite set of forbidden subgraphs
\cite{HS-GatexForbSubg:23}.
\gatex graphs that are explained by labeled trees $(T,t)$
are precisely the cographs and, therefore, those graphs
that do not contain induced $P_4$s \cite{Corneil:81}.

\paragraph{\bf Modular Decomposition (MD).}
A \emph{module} $M$ of a graph $G=(X,E)$ is a subset $M\subseteq V(G) = X$ such
that for all $x,y\in M$ it holds that $N_G(x)\setminus M = N_G(y)\setminus M$,
where $N_G(x)$ is the set of all vertices of $X$ that are adjacent to $x$ in
$G$. A module $M$ of $G$ is \emph{strong} if $M$ does not \emph{overlap} with
any other module of $G$, that is, $M\cap M' \in \{M, M', \emptyset\}$ for all
modules $M'$ of $G$. The set of strong modules $\MDstrong(G)\subseteq \MD(G)$ is
uniquely determined \cite{HSW:16,EHMS:94} and forms a hierarchy which gives rise
to a unique tree representation $\MDT_G$ of $G$, known as the \emph{modular
decomposition tree} (\emph{MDT}) of $G$. Uniqueness and the hierarchical
structure of $\MDstrong(G)$ implies that there is a unique partition $\Mmax(G) =
\{M_1,\dots, M_k\}$ of $X$ into inclusion-maximal strong modules $M_j\ne X$ of
$G$ \cite{ER1:90,ER2:90}. 

Similar as for galled-trees, one can equip $\MDT_G$ with a vertex-labeling $t_G$
such that, for $M\in \MDstrong(G) = V(\MDT_G)$, we have $t_G(M)=\odot$ if
$|M|=1$; $t_G(M)=0$ if $|M|>1$ and $G[M]$ is disconnected; $t_G(M)=1$ if $|M|>1$
and $G[M]$ is connected but $\overline G[M]$ is disconnected; $t_G(M) =
\mathit{prime}$ in all other cases. Strong modules of $G$ are called
\emph{series}, \emph{parallel} and \emph{prime} if $t_G(M)=1$, $t_G(M)=0$ and
$t_G(M) = \mathit{prime}$, respectively. Efficient linear algorithms to compute
$(\MDT_G,t)$ have been proposed e.g.\ in \cite{DGC:01,CS:99,TCHP:08}. The
\emph{quotient graph} $G/\Mmax(G)$ has $\Mmax(G)$ as its vertex set and edges
$\{M_i,M_j\}\in E(G/\Mmax(G))$ if and only if there are $x\in M_i$ and $y\in
M_j$ that are adjacent in $G$. As argued in \cite{HP:10}, this quotient
graph is well-defined.

\paragraph{\bf From Modular Decomposition Trees to Galled-trees.}
Galled-trees that explain a given \gatex graph $G$ can be obtained from the
modular decomposition trees $(\MDT_G,t_G)$ by replacing its prime vertices
locally by simple rooted cycles. To this end, we first compute for prime
vertices $v$ and the corresponding prime modules $M = L(\MDT_G(v))$ the quotient
$H = G[M]/\Mmax(G[M])$ which can be explained by  a strong elementary
quasi-discriminating galled-tree $(N_v,t_v)$ (cf.\ \cite[Thm.\ 6.10]{HS:22}). We
then use the rooted cycles in $(N_v,t_v)$ to replace $v$ in $(\MDT_G,t_G)$, see
Figure~\ref{fig-algexpl} for an illustrative example. The latter is formalized
as follows.

\begin{definition}[prime-vertex replacement (pvr) networks]
  \label{def:pvr}
    Let $G$ be a \gatex graph and $\mathcal{P}$ be the set of all prime vertices in
	$(\MDT_G,t_G)$. 
	 A \emph{prime-vertex replacement} (\emph{pvr}) networks $(N,	
	t)$ of $G$ (or equivalently, of $(\MDT_G,t_G)$) is obtained by the following procedure:
\begin{enumerate}
\item For all $v\in \mathcal{P}$, let $(N_v,t_v)$ be a
	  strong quasi-discriminating elementary galled-tree with root $v$
	  that explains $G[M]/\Mmax(G[M])$ with $M = L(\MDT_G(v))$.			
  \label{step:Gv} \smallskip
\item For all $v\in \mathcal{P}$, remove all edges $(v,u)$ with
  		$u\in \child_{\MDT_G}(v)$ from $\MDT_G$ to obtain the forest
  		$(T',t_G)$ and \label{step:T'}  
		add $N_v$ to $T'$ by identifying the root
  	of $N_v$ with $v$ in $T'$ and each leaf $M'$ of $N_v$ with the
  	corresponding child $u\in \child_{\MDT_G}(v)$ for which $M' = L(\MDT_G(u))$.  \smallskip

\noindent
 \emph{This results in the \emph{pvr-network $N$ of $G$}.}\smallskip
\item \label{step:color} 
 Define the labeling $t\colon V(N)\to \{0,1,\odot\}$ by putting, for
  all $w\in V(N)$,
  \begin{equation*}
    t(w) =
    \begin{cases} 
      t_G(v) &\mbox{if } v\in V(\MDT_G)\setminus \mathcal P \\
      	t_v(w) &\mbox{if } w \in V(N_v)\setminus X \text{ for some } v\in \mathcal P
    \end{cases}
  \end{equation*}
\end{enumerate}
\end{definition}

\begin{figure}[t]
	\begin{center}
			\includegraphics[width=.7\textwidth]{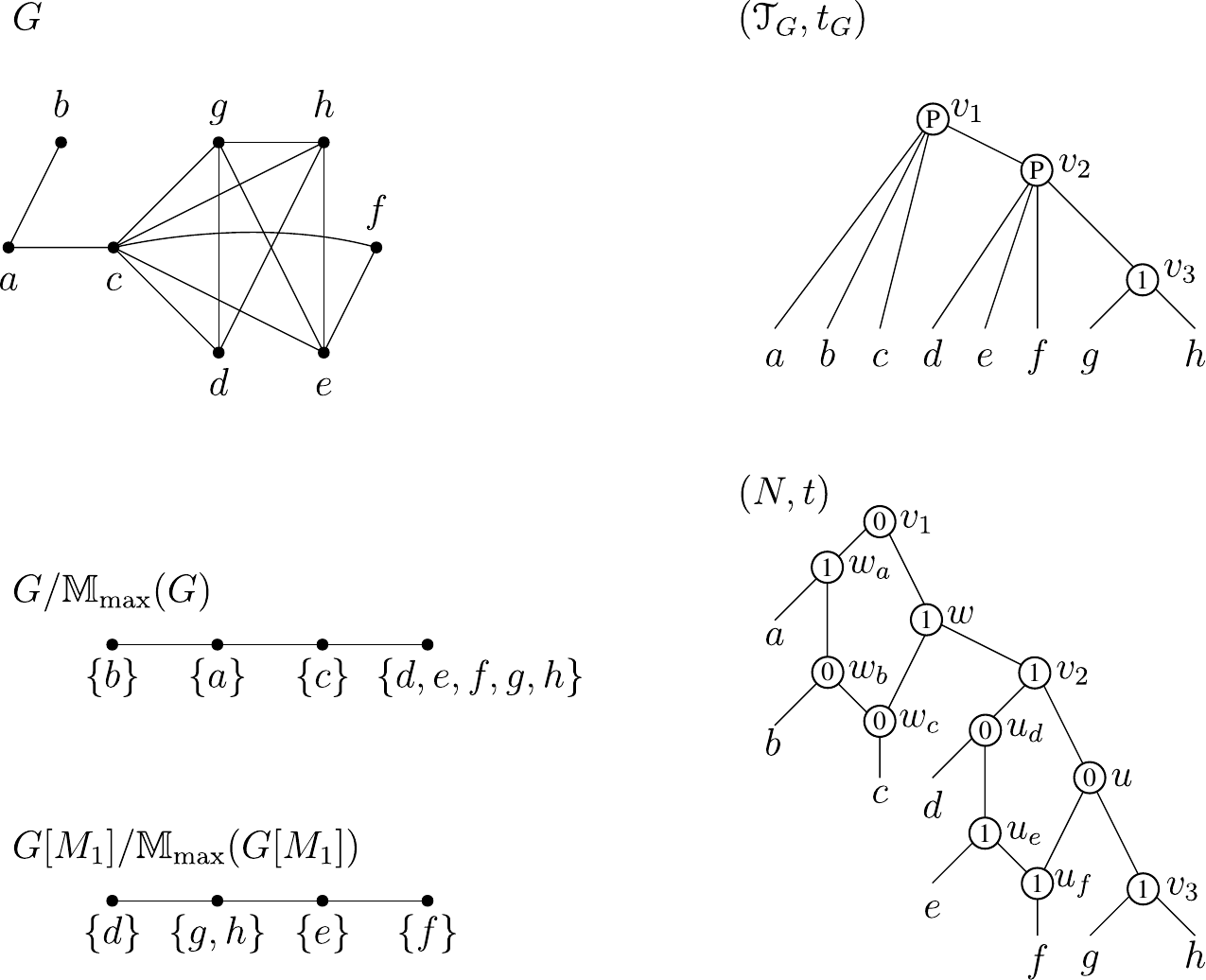}
	\end{center}
	\caption{Shown is a \gatex graph $G=(V,E)$ (top left) together with its 
	modular decomposition tree $(\MDT_G,t_G)$ (top right) and a
	pvr-network $(N,t)$ (bottom right) that explains $G$. 
	The graph $G$ has as strong modules the singletons $\{x\}$, $x\in V = \{a,b,c,\dots,g\}$, 
	the entire vertex set $V$ and the sets $M_1 =\{d,e,f,g,h\}$ and $M_2=\{g,h\}$. 
	Each vertex $w$ in $\MDT_G$ represent the strong module $L(\MDT_G(w))$. 
	The graph $G$ has two
	prime modules, namely $M_1 = L(\MDT_G(v_2))$ and 
	$V = L(\MDT_G(v_1))$. The respective quotient graphs $H_1 \coloneqq G/\Mmax(G)$ and 
	 $H_2 \coloneqq G[M_1]/\Mmax(G[M_1])$ are shown bottom left. The pvr-network  $(N,t)$ is a galled-tree that is obtained
	 from $(\MDT_G,t_G)$ by locally replacing the vertex $v_i$ by
	 the strong quasi-discriminating elementary galled-tree  $(N_{v_i},t_{v_i})$ 
	 that explains $H_i$, $i\in \{1,2\}$ (cf.\ Def.~\ref{def:pvr}).
	 %	Shown is a \gatex graph $G$ (right), its modular decomposition tree
%	         $(\MDT_G,t_G)$ (left) and a pvr network $(N,t)$ that explains $G$
%	         (middle). The inner vertices of $(\MDT_G,t_G)$ and $(N,t)$ have
%	         label ``1'', ``0'' and ``P'' (prime). One easily observes that
%	         $(N,t)$ is obtained from $(\MDT_G,t_G)$ by replacing the prime
%	         vertices by rooted 0/1-labeled cycles. Using Alg.\ \ref{alg:sigma},
%	         we obtain a perfect order $\zeta=dabcegfijhkl$ of $G$ (see text
%	         for details). The graph $G$ is colored w.r.t.\ the greedy coloring
%	         based on the order $\zeta$ (the order of colors is shown in the
%	         figure.) Since $G[\{g,h,i,k\}]\simeq K_4$, this coloring is
%	         optimal, i.e., it uses $\chi(G)=4$ colors. 
			 }
\label{fig:example}
\end{figure}

Note that the leaves of the pvr-network $N$ of $G$ are the singletons $\{x\}$, $x \in V(G)$. 
In the remainder of this paper, we will always implicitly identify each singleton with its unique elements. 
In other words, we will always assume that the leaf set of $\MDT_G$ as well as of
pvr-network $N$ of $G$ is $V(G)$. By construction, we have $V(\MDT_G) \subseteq V(N)$ given that $N$ is the pvr-network of $G$.
More precisely, $V(\MDT_G)$ is precisely the set of vertices $v$ of $N$ such that either $v$ does not belong to a cycle $C$ of $N$, or $v=\rho_C$ for some cycle $C$ of $N$. 
In addition, we have
\begin{observation}\label{obs:ancest}
For a vertex $v$ in in the pvr-network $N$ of $G$, the following holds:
\begin{itemize}
\item[(i)] If $v$ does not belong to any cycle in $N$, then all children of $v$ in $N$ are children of $v$ in $\MDT_G$.
\item[(ii)] If there exists a cycle $C$ of $N$ such that $v \in V(C) \setminus \{\rho_C\}$, then $v$ has a unique child $w$ in $V(N) \setminus V(C)$, and $w$ is a child of $\rho_C$ in $\MDT_C$.
\end{itemize}
\end{observation}

The construction of a pvr-network for a \gatex graph is well-defined and can be done in linear-time,
cf.\ \cite[Alg.\ 4 \& Thm.\ 9.4]{HS:22}. By \cite[Prop.\ 7.4 \& 8.3]{HS:22}, 
a pvr-network $(N,t)$ of a \gatex graph $G$ is a galled-tree that explains $G$.
Moreover, there is a 1:1 correspondence between cycles $C$ in $N$ and 
prime modules $M$ of $G$. By the latter result, we can define $C_M$ as the unique
cycle in $N$ corresponding to prime module $M$. 
For later reference, we summarize now a couple of results that are easy to verify or that have been established in \cite{HS:22}.

%\begin{observation}\label{obs:properties_C_pvr}
%Let $v$ be a prime vertex associated with the prime module $M_v=L(\MDT_G(v))$ module 
%and let $C \coloneqq C_{M_v}$. Since we used strong elementary networks for the replacement
%of $v$, one easily verifies that:
%\begin{itemize}
%\item $C$ has a unique root $\rho_C$ and a unique hybrid-vertex $\eta_C$. 
%\item $\eta_C$ has precisely one child and precisely two parents. 
%\item All vertices $v\neq \eta_C$ in $C$ have two children and one parent. 

%	 In particular, 
%	  all vertices $v\neq \eta_C,\rho_C$ in $C$ 	
%	   have one child $v_1$ located in $C$
%	  and one child $v_2$ that is not located in $C$
%	  and these children satisfy  $L(N(v_1))\cap L(N(v_2))=\emptyset$
%	  and it holds that $\lca_N(x,y)=v$ for all $x\in L(N(v_1))$ and $y\in L(N(v_2))$. 
%		 
%	  Both children $u'$ and $u''$ of 
%	  $\rho_C$ are located in $C$ and satisfy  $L(N(u'))\cap L(N(u''))=L(N(\eta_C))$. 
%	  Moreover, $L(N(\eta_C))\cap L(N(v_2)) = \emptyset$ for the
%	   child $v_2$   of  $v\neq \eta_C,\rho_C$ that is not located in $C$.
%\end{itemize}
%\end{observation}

\begin{observation}\label{obs:properties_C_pvr} 
Let $(N,t)$ be a pvr-network of a \gatex graph $G$. Then, 
\begin{itemize}
\item $(N,t)$ is a galled-tree that explains $G$ \cite[Prop.\ 7.4]{HS:22}.
\item There is a 1:1 correspondence between the cycle $C$ in $N$ and 
	  prime modules $M$ of $G$ \cite[Prop.\ 8.3]{HS:22}.
	  
	  Hence, we can define 
      $C_M$ as the unique cycle in $N$ corresponding to prime module $M$. 
      
      Moreover, 
 $G_1(M), G_2(M)\subseteq G[M]$ will denote the subgraphs induced by leaf-descendants of
the vertices in $P^1(C_M)-\rho_{C_M}$ and $P^2(C_M)-\rho_{C_M}$, respectively.

\end{itemize}
Moreover, let $v$ be a prime vertex associated with the prime module $M_v=L(\MDT_G(v))$ module 
and let $C \coloneqq C_{M_v}$. Since we used strong elementary networks for the replacement
of $v$, one easily verifies that:
\begin{itemize}
\item $C$ has a unique root $\rho_C$ and a unique hybrid-vertex $\eta_C$. 
\item $\eta_C$ has precisely one child and precisely two parents. 
\item All vertices $v\neq \eta_C$ have two children and one parent. 

	 In particular, 
	  all vertices $v\neq \eta_C,\rho_C$
	   have one child $u'$ located in $C$
	  and one child $u''$ that is not located in $C$
	  and these children satisfy  $L(N(u'))\cap L(N(u''))=\emptyset$
	  and it holds that $\lca_N(x,y)=w$ for all $x\in L(N(u'))$ and $y\in L(N(u''))$.

	  Both children $u'$ and $u''$ of 
	  $\rho_C$ are located in $C$ and satisfy  $L(N(u'))\cap L(N(u''))=L(N(\eta_C))$. 
	  Moreover, $L(N(\eta_C))\cap L(N(v_2)) = \emptyset$ for the
	   child $v_2$   of  $v\neq \eta_C,\rho_C$ that is not located in $C$.

%\item  For all $w\in C\neq \{\eta_C,\rho_C\}$ and its two children $u',u''$
%       we have $\lca_N(x,y)=w$ for all $x\in L(N(u'))$ and $y\in L(N(u''))$. 
%       Moreover, if  $v,w\in C$ are $\preceq_N$-incomparble, then 
%       $\lca_N(x,y)=\rho_C$ for all $x\in L(N(v))$ and $y\in L(N(w))$. 
\end{itemize}

\end{observation}

\section{Perfect orderings and optimal colorings}

In this section, we provide  linear-time algorithms to compute
the chromatic number $\chi(G)$ and an optimal coloring of a given \gatex graph $G$. For this
purpose, we show first how to employ the structure of labeled galled-trees
$(N,t)$ to determine a perfect ordering of \gatex graphs in linear time (cf.\
Alg.\ \ref{alg:sigma}). To this end, we provide the following result for
later reference.

\begin{lemma}\label{lem:abcd}
Let $P=a-b-c-d$ be an induced $P_4$ in a \gatex graph $G$ and $(N,t)$ a pvr-network
that explains $G$. Moreover, let $M$ be the inclusion-minimal strong module of $G$
that contains $V(P)$, i.e., $V(P)\subseteq M$ and there is no strong module $M'$ of $G$ 
 that satisfies $V(P) \subseteq M' \subsetneq M$. Then, $M$ is a prime module of $G$. 
 Moreover, in the unique cycle  $C_M$ in $N$ that corresponds to $M$, 
 there are vertices $u_a,u_b,u_c,u_d \in V(C_M)$ that 
 satisfy the following conditions:
\begin{enumerate}
\item  For $x\in \{a,b,c,d\}$ it holds that $x\in L(N(u'_x))$
       where $u'_x$ is the unique child of $u_x$ that is not located in $C_M$.
\item  The vertices $u_a,u_b,u_c,u_d$ are pairwise distinct.
\item  The vertices $u_a,u_b,u_c,u_d$ do not all belong to the same side of $C_M$.
\item  One of $u_a,u_b,u_c,u_d$ coincides with the unique hybrid $\eta_{C_M}$ of $C_M$.
\end{enumerate}
\end{lemma}
\begin{proof}
Let $P=a-b-c-d$ be an induced $P_4$ in a \gatex graph $G$ and $(N,t)$ a
pvr-network that explains $G$. Put $Y \coloneqq \{a,b,c,d\}$. Moreover, let $M$
be the inclusion-minimal strong module of $G$ that contains $V(P)$.

We show first that $M$ is a prime module of $G$. By definition, we must
show that $G[M]$ and $\overline G[M]$ are connected. Assume that $P=a-b-c-d$ and
put $Y\coloneqq V(P)$. Observe first that $G[Y] = P$ and $\overline G[Y] =
c-a-d-b$. Hence, both $G[Y]$ and $\overline G[Y] $ are connected. Assume, for
contradiction, that $G[M]$ is disconnected. In this case, $P$ belongs to some
connected component $H$ of $G[M]$.
We show that, in this case, $V(H)$ must be a strong module of $G[M]$. 
Clearly, $V(H)$ is a module of $M$. Assume, for contradiction, that
$V(H)$ is not strong. Hence, it overlaps with some module $M'$ of $G[M]$ and, therefore, 
$V(H)\cap M'\neq \emptyset$, $V(H)\setminus M'\neq \emptyset$
and $M'\setminus V(H)\neq \emptyset$. In particular, since 
$H$ is connected, there is a vertex $x\in V(H)\cap M'$ that is
adjacent to some vertex $y\in V(H)\setminus M'$. However, since 
$H$ is a connected component, none of the vertices 
$z\in M'\setminus V(H)$ can be adjacent to $y$. Hence, 
$M'$ is a not a module; a contradiction. Thus, $V(H)$ is a strong module of
$G[M]$ and, by \cite[Lemma 3.1]{Hellmuth:20b}, $V(H)$
is a strong module of $G$; a contradiction to the choice of $M$. Thus, $G[M]$ is
connected. By similar arguments and since $\overline G[Y]= c-a-d-b$, $\overline
G[M]$ must be connected as well. Consequently, $M$ is a prime module of $G$.

Since $M$ is a prime module of $G$, there is a unique cycle $C_M$ in $N$
corresponding to $M$.
To recall, $L_w = L(N(w))$ for $w\in
V(N)$. For a vertex $w\in V(\MDT_G)$, we denote with $M_w\coloneqq L(\MDT_G(w))$
the module of $G$ ``associated'' with $w$. 
For all vertices $u \in V(C) \setminus \{\rho_C\}$, we put  $L'_u=L_{u'}$, where $u'$ is the unique child of
$u$ that is not in $V(C)$. By
Obs.~\ref{obs:properties_C_pvr}, the sets $L'_u, u \in V(C) \setminus
\{\rho_C\}$ are pairwise disjoint strong modules of $G$. In particular, for all
$x \in M$, there exists a unique vertex $u_x \in V(C) \setminus \{\rho_C\}$ such
that $x \in L'_{u_x}$. Consequently, Condition (1) is satisfied.

We show now that the vertices $u_a, u_b, u_c$ and $u_d$ are pairwise distinct. If
$u_a=u_b=u_c=u_d$, then $Y \subseteq L'_{u_a}$. Since $L'_{u_a}$ is a strong
module of $G$ satisfying $L'_{u_a} \subsetneq M$, this contradicts the choice of
$M$. If exactly three of $u_a, u_b, u_c, u_d$ are equal, then there exists $u
\in V(C) \setminus \{\rho_C\}$ and $x \in Y$ such that $Y \setminus \{x\}
\subseteq L'_{u}$ and $x\notin L'_u$. This and the fact that $L'_{u}$ is a
module of $G$ implies that either all or none of of the vertices in $L'_u$
(and thus, of $Y \setminus \{x\}$) are adjacent to $x$. Consequently, $x$ has degree $0$ or $3$ in $G[Y]$, a contradiction since
$G[Y]=P$. Finally, if two of $u_a, u_b,
u_c, u_d$ are equal and distinct from the other two, then there exists $u \in
V(C) \setminus \{\rho_C\}$ and $x,y \in Y$ distinct such that $Y \cap
L'_u=\{x,y\}$. Since $L'_u$ is a module, then for all $z \in Y \setminus
\{x,y\}$, $\{x,z\}$ is an edge of $G[Y]$ if and only if $\{y,z\}$ is an edge of
$G[Y]$. However, since $G[Y]=P$, there is no
pair $\{x,y\}$ of elements of $Y$ satisfying this property. Therefore, the
vertices $u_a, u_b, u_c$ and $u_d$ are pairwise distinct and
Condition (2) is satisfied. 

This in particular implies that, for $x,y \in Y$
distinct, $\lca_N(x,y) \in \{u_x,u_y, \rho_C\}$. More specifically, we
$\lca_N(x,y)=u_x$ if $u_y\prec_N u_x$, $\lca_N(x,y)=u_y$ if $u_x\prec_N u_y$, and $\lca_N(x,y)=\rho_C$ if $u_x$ and
$u_y$ are $\preceq_N$-incomparable. 
Assume, for contradiction, that the vertices $u_a, u_b, u_c$ and $u_d$ all belong to
the same side of $C$. In this case, there is a vertex $x \in Y$ such that
$u_x$ is an ancestor of $u_a, u_b, u_c$ and $u_d$ in $N$. In view of the above,
$\lca_N(x,y)=u_x$ for all $y \in Y \setminus \{x\}$. Hence, $x$
has degree $0$ in $G[Y]$ if $t(u_x)=0$, and degree $3$ if $t(u_x)=1$. Since
$G[Y]=P$, none of these cases can occur. Hence, Condition (3) is satisfied.

We now show that one of $u_a, u_b, u_c$ and $u_d$ coincides with $\eta\coloneqq \eta_{C}$. Assume,
for contradiction, that this is not the case. Then two situations may occur:
exactly three of $u_a, u_b, u_c$ and $u_d$ belong to the same side of $C$, or
two of $u_a, u_b, u_c$ and $u_d$ belong to one side of $C$, and the other two
belong to the other side. Suppose first that there exists $x \in Y$ such that
$u_x$ is the only vertex on its side of $C$. Then we have $\lca_N(x,y)=\rho_C$
for all $y \in Y \setminus \{x\}$. Hence, $x$ has degree $0$ in $G[Y]$ if
$t(\rho_C)=0$, and degree $3$ if $t(\rho_C)=1$. Since $G[Y]=P$, both cases are impossible. Suppose now that there exists $x,y \in Y$ distinct
such that $u_x$ and $u_y$ belong to one side of $C$. In particular, for all $z \in Y \setminus \{x,y\}$, we have $\lca_N(x,z)=\lca_N(y,z)=\rho_C$. It follows that $G[Y]$ is disconnected if $t(\rho_C)=0$ and $\overline G[Y]$ is disconnected if $t(\rho_C)=1$. Since $G[Y]=P$, both 
cases are impossible. Hence, one of $u_a, u_b, u_c$ and $u_d$ coincides with $\eta$ and Condition (4) is satisfied.
\end{proof}

\begin{algorithm}[t]
  \small
  \caption{\texttt{Perfect ordering of \gatex graphs  $G$}}
  \label{alg:sigma}
  \begin{algorithmic}[1]
    \Require  A \gatex graph $G=(V,E)$
    \Ensure   A perfect ordering $\zeta$ of the vertices of $V(G)$  

    \State Construct  $(\MDT_G,t_G)$ and pvr-network $(N,t)$ of $G$ \label{l:MDT-pvr2}

    \State Initialize $\zeta(v)\coloneqq v$ for all leaves $v$ in $\MDT_G$ \label{l:sigma-leaves}
    \ForAll{$v\in V(\MDT_G)\setminus{L(\MDT_G)}$ in postorder} \label{l:forV2}
    
    	\If{$t_G(v) \in \{0,1\}$} \label{l:v-nonprime}
    		       
    		 \State Put $\zeta(v)\coloneqq\zeta(v_1) \ldots \zeta(v_k)$ arbitrarily for the $k=|\child_{\MDT_G}(v)|$ 
    		 	children $v_1, \ldots, v_k$ of $v$ in $\MDT_G$ \label{l:sigmanprime}
       
    	\Else  \Comment{$t_G(v)=\mathrm{prime}$} \label{l:v-prime2}

    	 	\State Let $C$ be the unique cycle in $N$ with root $\rho_C=v$. \label{l:cycle}
    	 	\State For all vertices $w \in V(C)\setminus \{\rho_C\}$, put $\zeta(w)\coloneqq \zeta(w')$, where $w'$ is 
    	 		   the unique child of $w$ in $N$ that is not a vertex of $C$. \label{l:sigmaC}
		\State Put $\zeta^*(v)\coloneqq\zeta(v_1) \ldots \zeta(v_k)$ arbitrarily for the $k=|V(C)|-2$ 
			vertices $v_1, \ldots, v_k$ in $V(C)\setminus\{\rho_C,\eta_C\}$ \label{l:zeta-star}
    	 	\If{$t(\rho_C)=0$}\label{l:rho-check-start}
    	 		\State$\zeta(v)\coloneqq\zeta(\eta_C)\zeta^*(v)$ \label{l:sigmap0}
    	 	\Else \Comment{$t(\rho_C)=1$}
    	 		\State $\zeta(v)\coloneqq\zeta^*(v)\zeta(\eta_C)$ \label{l:sigmap1}
    	 	\EndIf \label{l:rho-check-end}
    	 \EndIf
	\EndFor \label{l:forV-end}
	\State \Return $\zeta(v)$
  \end{algorithmic}
\end{algorithm}

\begin{figure}[t]
	\begin{center}
			\includegraphics[width=.6\textwidth]{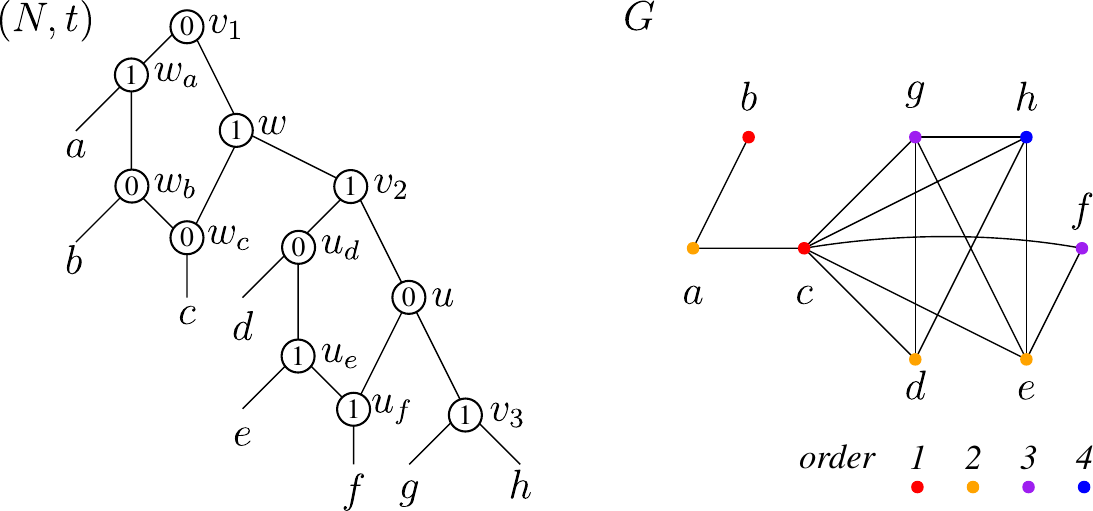}
	\end{center}
	\caption{Left a  galled-tree $(N,t)$ that explains the \gatex graph $G$ on the right. 
					In addition, $G$ is equipped with a vertex coloring that is obtained
					with a greedy coloring based on the perfect order $cabdeghf$ computed
					with Algorithm~\ref{alg:sigma}. Since $G[c,e,g,h]$ is a complete graph on four vertices, this coloring is optimal. 
					see explanations in Example \ref{exmpl:alg-order} for further details.
}
\label{fig-algexpl}
\end{figure}

As we shall see, Algorithm~\ref{alg:sigma} can be used to compute a perfect order in \gatex graphs in linear-time.
Before studying Algorithm~\ref{alg:sigma} in detail, we illustrate this
algorithm on the example shown in Figure~\ref{fig-algexpl}. 

\begin{example}\label{exmpl:alg-order}
We exemplify here the main steps of Algorithm~\ref{alg:sigma} using as input the \gatex graph $G$ as
shown in Fig.~\ref{fig-algexpl}. We first compute the modular decomposition tree $(\MDT_G,t_G)$
(as shown in Fig.~\ref{fig:example}) and the shown pvr-network $(N,t)$ that explains $G$
(Line \ref{l:MDT-pvr2}). For all leaves $v$ of $\MDT_G$ (and thus, of $N$), we initialize the
perfect order $\zeta(v) = v$ of the induced subgraph $G[\{v\}]$ (Line~\ref{l:sigma-leaves}). We then
traverse the vertices $\MDT_G$ that are not leaves in postorder and thus obtain the order
$v_3,v_2,v_1$ in which the vertices are visited (Line~\ref{l:forV2}). Note that postorder-traversal
ensures that all children of a given vertex $v$ in $\MDT_G$ are visited before this vertex $v$ is
processed. Since $v_3$ is a non-prime vertex of $\MDT_G$ (Line~\ref{l:v-nonprime}), we can
choose one of the orders $\zeta(g)\zeta(h)$ or $\zeta(h)\zeta(g)$ (Line~\ref{l:sigmanprime}) and
decide, in this example, to put $\zeta(v_3)=\zeta(g)\zeta(h)=gh$. We proceed with vertex $v_2$
which is a prime vertex in $\MDT_G$. We now consider the cycle $C$ with root $\rho_C=v_2$
(Line~\ref{l:cycle}). This cycle $C$ refers to the subgraph in $N$ induced by $v_2, u,
u_d,u_e,u_f$. In Line~\ref{l:sigmaC}, we put $\zeta(u_x)=\zeta(x)=x$ for each $x \in \{d,e,f\}$ and
$\zeta(u)=\zeta(v_3)=gh$. In Line \ref{l:zeta-star} we can choose an arbitrary ordering
$\zeta^*(v_2)$ and decide, in this example, for $\zeta^*(v_2) = \zeta(u_d)\zeta(u_e)\zeta(u) =
degh$. Since $u_f=\eta_C$ and $t(v_2)=1$, we put $\zeta(v_2)=\zeta^*(v_2)\zeta(\eta_C)=deghf$
(Line~\ref{l:sigmap1}). Finally, the prime vertex $v_1$ is processed. We consider now the cycle $C$
with root $\rho_C=v_1$ that is induced by $v_1,w,w_a,w_b,w_c$. (Line~\ref{l:cycle}). In
Line~\ref{l:sigmaC}, we put $\zeta(w_x)=\zeta(x)=x$ for each $x \in \{a,b,c\}$ and
$\zeta(w)=\zeta(v_2)=deghf$. In Line \ref{l:zeta-star} we can choose an arbitrary ordering
$\zeta^*(v_1)$ in Line \ref{l:zeta-star} and decide, in this example, for $\zeta^*(v_1) =
\zeta(w_a)\zeta(w_b)\zeta(w) = abdeghf$. Finally, since $t(v_1)=0$ and $\eta_C=w_c$, we put
$\zeta(v_1)=\zeta(\eta_C)\zeta^*(v_1)=cabdeghf$ (Line~\ref{l:sigmap0}). Since $v_1$ is the root of
$\MDT_G$, the algorithm stops there, and returns the ordering $\zeta=\zeta(v_1)=cabdeghf$. As we
shall show in Prop.\ \ref{prop:AlgPerfectOrder}, this ordering is a perfect ordering It is now
an easy task to verify that the vertex coloring of $G$ as shown in Fig.~\ref{fig-algexpl} can be
obained by a greedy coloring taking the perfect order $\zeta=cabdeghf$ and the order of colors as
shown in Fig.~\ref{fig-algexpl} (bottom right) into account.
\end{example}

\begin{proposition}\label{prop:AlgPerfectOrder}
Algorithm \ref{alg:sigma} determines a perfect ordering of \gatex graphs.
\end{proposition}
\begin{proof}
Let $G=(V,E)$ be a \gatex graph that serves as input for Alg.~\ref{alg:sigma}.
We first compute $(\MDT_G,t_G)$ and a pvr-network $(N,t)$ of $G$ (Line
\ref{l:MDT-pvr2}). 
In this proof, we put $L_w \coloneqq L(N(w))$ for $w\in
V(N)$. 
Let $\zeta(w)$ be the ordering
computed with Alg.~\ref{alg:sigma} for the subgraph $G[L_w]$ induced by the
vertices in $L_w$.
We then initialize $\zeta(v) = v$ for all leaves $v$ in $\MDT_G$ (Line
\ref{l:sigma-leaves}). Clearly, $\zeta(v)$ is a perfect ordering of
$G[\{v\}]$. We then continue to traverse the remaining
vertices in $\MDT_G$ in postorder. This ensures that, whenever we reach a vertex
$v$ in $\MDT_G$, all its children have been processed and thus, 
that $\zeta(v)$ is well-defined in each step.

To verify that the ordering $\zeta$ returned by Alg.~\ref{alg:sigma} is a
perfect order of $G$, we must show that $\zeta$ does not contain any
obstructions w.r.t. $G$ (cf.\ Prop.~\ref{prop:perfect-order}). If $G$ does not
contain any induced $P_4$, then any ordering is perfect. Thus, assume that $G$
contains an induced $P_4$, say $P=a-b-c-d$. Put $Y=\{a,b,c,d\}$.

We first remark that Alg.~\ref{alg:sigma} builds $\zeta$ by successively
concatenating sub-orderings of the form $\zeta(w)$, $w \in V(\MDT(G))$. In
particular $\zeta_{|Y}=\zeta(w)_{|Y}$ holds for all $w \in V(\MDT_G)$ for which
$Y \subseteq M_w$ where $M_w\coloneqq L(\MDT_G(w))$. Let $M$ be the
inclusion-minimal strong module of $G$ that contains $Y$. 
By Lemma \ref{lem:abcd}, $M$ is a prime module of $G$. 
Hence, there is   the unique cycle $C\coloneqq
C_M$ in $N$ corresponding to $M$. For all vertices $u \in
V(C) \setminus \{\rho_C\}$, we denote with $u'$ is the unique child
of $u$ that is not in $V(C)$. By Lemma \ref{lem:abcd}, there are four vertices
$u_a,u_b,u_c,u_c \in V(C)$ that satisfy the Condition (1) - (4). Hence, for
$x\in \{a,b,c,d\}$ it holds that $x\in L_{u'_x}$. Moreover, the vertices
$u_a,u_b,u_c,u_c$ are pairwise distinct, do not all belong to the same side of
$C_M$ and one of $u_a,u_b,u_c,u_c$ coincides with the unique hybrid
$\eta\coloneqq \eta_{C}$ of $C$. The latter arguments, in particular, allow us to denote by
$P^-$ (resp., $P^+$) the side of $C$ such that the set $V(P^-) \setminus
\{\eta\}$ (resp., $V(P^+) \setminus \{\eta\}$) contains one (resp., two) of
$u_a, u_b, u_c$ and $u_d$. In the following, let $v$ be the prime vertex 
in $\MDT_G$ with $L(\MDT_G(v)) = M$. We now distinguish between two cases: (1)
$t(\rho_C)=0$ and (2) $t(\rho_C)=1$.

Case (1): $t(\rho_C)=0$. Let $x \in Y$ be the vertex such that 
$u_x \in V(P^-) \setminus \{\eta\}$. Then, for all $y \in Y \setminus \{x\}$ with
$u_y \in V(P^+) \setminus \{\eta\}$, we have $\lca_N(x,y)=\rho_C$ and thus,
$x$ and $y$ are not joined by an edge in $G[Y]$. 
In particular, $x$ has degree at most one in $G[Y]$. Since $G[Y]=P$,
it follows that $x$ has degree exactly one in $G[Y]$, and
that the unique vertex $z \in Y$ adjacent to $x$ in $N$ satisfies $u_z=\eta$. 
Due to the ``symmetry'' of $G[Y]=P = a-b-c-d$, 
we can assume w.l.o.g.\ that $x=a$ and thus, $z=b$.
By construction of $\zeta(v)$ in Line~\ref{l:sigmap0},
we have $\zeta(v) = \zeta(\eta)\zeta^*(v)$. Since vertex $b$
appears in the order $\zeta(\eta)$ and vertex $a$ appears in the
order $\zeta^*(v)$, we have in the final order
$\zeta$ of $G$ always $b<a$. In this case, $P$ does not yield
an obstruction of $\zeta$.

Case (2): $t(\rho_C)=1$. Let $x \in Y$ be the vertex such that $u_x
\in V(P^-) \setminus \{\eta\}$. Then for all $y \in Y \setminus \{x\}$ such that
$u_y \in V(P^+) \setminus \{\eta\}$, we have $\lca_N(x,y)=\rho_C$ and thus,
$x$ and $y$ are joined by an edge in $G[Y]$. In particular, $x$ has degree at least two in $G[Y]$. 
Since $G[Y]=P$, it follows that $x$ has degree exactly two in $G[Y]$, and
that the unique vertex $z \in Y$ that is \emph{not} adjacent to $x$ in $N$ satisfies
$u_z=\eta$. Again, by ``symmetry'' of $G[Y]=P = a-b-c-d$, 
we can assume w.l.o.g.\ that $x=c$ and thus, $z=a$.
Now consider the unique vertex $b$ that is adjacent to $a$ in $G[Y]$. 
By assumption, $b\in V(P^+) \setminus \{\eta\}$. 
Furthermore, by construction of $\zeta(v)$ in Line~\ref{l:sigmap1},
we have $\zeta(v) = \zeta^*(v)\zeta(\eta)$.
Since vertex $a$ appears in the order $\zeta(\eta)$ and vertex $b$ appears in  the
order $\zeta^*(v)$, we have in the final order
$\zeta$ of $G$ always $b<a$. In this case, $P$ does not yield
an obstruction of $\zeta$.

In summary, the ordering $\zeta$ returned by Alg.~\ref{alg:sigma} does not contain any
obstructions w.r.t. $G$. By Prop.~\ref{prop:perfect-order}, $\zeta$ is a perfect order
of $G$.
\end{proof}

\begin{proposition}\label{prop:AlgPerfectOrder-runtime}
Algorithm \ref{alg:sigma} 
can be implemented to run in $O(|V|+|E|)$ time
where $G=(V,E)$ is the input \gatex graph. 
\end{proposition}
\begin{proof}
 We show now that Algorithm \ref{alg:sigma} can be implemented to run in
 $O(|V|+|E|)$ time for a given \gatex graph $G=(V,E)$. The modular decomposition
 tree $(\MDT_G,t_G)$ can be computed in $O(|V|+|E|)$ time \cite{HP:10}. By
 \cite[Thm.\ 9.4 and Alg.\ 4]{HS:22}, the pvr-network $(N,t)$ of $G$ can be
 computed within the same time complexity. Thus, Line \ref{l:MDT-pvr2} takes
 $O(|V|+|E|)$ time. Initializing $\zeta(v)\coloneqq v$ for all leaves $v$ (and
 thus, the vertices of $G$) in Line \ref{l:sigma-leaves} can be done in $O(|V|)$
 time. 
 
 We then traverse each of the $O(|V|)$ vertices in $(\MDT_G,t_G)$ in postorder.
 To compute the final perfect order, we consider an auxiliary directed graph $H$ that,
 initially, just consists of the vertices in $V$ and is edge-less. Whenever, we
 concatenate $\zeta'$ and $\zeta''$, we simply add an edge $(u,v)$ from the
 maximal element $u$ in $\zeta'$ to the minimal element $v$ in $\zeta''$ and
 define the minimal element of this now order $\zeta''' = \zeta'\zeta''$ as the
 minimal element of $\zeta'$ and the maximal element of $\zeta'''$ as the
 maximal element of $\zeta''$. Since we can keep track of these maximal and
 minimal elements (starting with $\zeta(v)\coloneqq v$ for all leaves $v$ and
 defining $v$ as the maximal and minimal element of $\zeta(v)$) in each of the
 steps, the concatenation of two orders $\zeta'$ and $\zeta''$ and updating the
 maximal and minimal of $\zeta''' = \zeta'\zeta''$ can be done in constant time.
 The final graph $H$ then consists of a single directed path that traverses each
 vertex in $V$. If $t_G(v)\in \{0,1\}$, then we pick an arbitrary ordering of
 the children of $v$ and define $\zeta(v) =\zeta(v_1) \ldots \zeta(v_k)$ by
 concatenating the orderings of its $k$ children $v_1,\dots,v_k$ (Line
 \ref{l:v-nonprime} - \ref{l:sigmanprime}). By the latter arguments, this task
 can be done in $O(|\child_{\MDT_G}(v)|)$ time for each non-prime vertex $v$.
 Otherwise, if $t_G(v) = \mathrm{prime}$, we consider the unique cycle $C$ in
 $N$ that satisfies $L(N(\rho_{C})) = L(\MDT_G(v))$ in Line \ref{l:cycle}. We
 note that we can keep track of $C$ and its correspondence to $v$ when
 constructing the pvr-network $(N,t)$ based on $(\MDT_G,t_G)$ and thus have
 constant-time access to these cycles $C$ in $N$. The assignment
 $\zeta(w)=\zeta(w')$ for all $w \in V(C)\setminus \{\rho_C\}$ can be done in
 $O(|V(C)|)$ time (Line \ref{l:sigmaC}). 
 By the latter arguments, construction of $\zeta^*(v)$ in Line \ref{l:zeta-star} can 
  be done in  $O(|V(C)|)$ time.
 Note that $O(|V(C)|) = O(|\child_{\MDT_G}(v)|)$, since the elementary galled-tree $N_v$ that is used
 to replace $v$ and the edges to its children in $\MDT_G$, contains $C$ and has
 $2\child_{\MDT_G}(v)+1$ edges and vertices. The tasks in Line
 \ref{l:rho-check-start}-\ref{l:rho-check-end} can be done in constant time.
 Hence, the time-complexity of the Lines \ref{l:v-prime2} to
 \ref{l:rho-check-end} is in $O(|\child_{\MDT_G}(v)|)$ for each prime vertex
 $v$.

 To obtain the overall time complexity of the \emph{for} loop starting in Line
 \ref{l:forV2}, observe that the degrees of vertices in $\MDT_G$ sum up to
 $2|E(\MDT_G)| = 2(|V(\MDT_G)|-1)$. By the latter arguments and by iterating
 over each vertex $v\in V(\MDT_G)\setminus L(\MDT_G)$, we obtain $\sum_{v\in
 V(\MDT_G)\setminus L(\MDT_G)} O(|\child_{\MDT_G}(v)|) = O(|V(\MDT_G)|) =
 O(|V|)$.

 Hence, the overall time-complexity of Algorithm \ref{alg:sigma} is dominated by
 the time-complexity to compute $(\MDT_G,t_G)$ and $(N,t)$ in Line
 \ref{l:MDT-pvr2} and is, therefore, in $O(|V|+|E|)$.
\end{proof}

As an immediate consequence of Prop.\  \ref{prop:AlgPerfectOrder} and \ref{prop:AlgPerfectOrder-runtime}, we obtain
\begin{theorem}\label{thm:perf-order-final}
Every \gatex graph is perfectly orderable and this ordering can be determined in linear-time. 
\end{theorem}

For a given graph $G=(V,E)$, a greedy coloring algorithm can be implemented
to run in $O(|V|+|E|)$ time, see e.g.\ \cite[Sec.~6.4]{TurauWeyer:2015}. 
This  together with Theorem \ref{thm:perf-order-final} implies
\begin{theorem}\label{thm:coloring}
The chromatic number $\chi(G)$ and an optimal coloring of a \gatex graph $G$   can be determined in linear-time. 
\end{theorem}

\section{Maximum cliques and independent sets}

To recall, a clique of a graph $G$ is an inclusion-maximal complete subgraph $G$
and  the maximum size of a clique of $G$ is denoted by $\omega(G)$. 
If $G$ is a \gatex graph, then it is explained by some labeled galled-tree
$(N,t)$ whose leaf set is $V(G)$.

Since \gatex graphs $G$ are perfect, their chromatic number $\chi(G)$ and the size $\omega(G)$
of a maximum clique coincide. This together with Theorem \ref{thm:coloring} implies

\begin{theorem}
The clique number $\omega(G)$ of a \gatex graph $G$  can be determined in linear-time. 
\end{theorem}

It is clear that for a given graph $G$ and integer $k = \omega(G)$, one can determine in
$O(|V(G)|^k)$ time a maximum clique by examining all $O(|V(G)|^k)$ subgraphs. Since $k = \omega(G)$
can be obtained in linear time for \gatex graphs, we, therefore, immediately obtain a
polynomial-time procedure to find maximum cliques in \gatex graphs.
In what follows, we show that maximum cliques
in \gatex graphs even  can be determined in linear time. 

To this end, we examine the structure and size of maximum-sized cliques 
induced by vertex set $L(N(v))$ in $G$ where $(N,t)$ is a galled-tree that explains $G$. 
In this context, it is important to take the labeling $t(v)$ of $v$ into account.

%\ref{obs:ancest}

%\TODO{list simple properties of pvr N: hybrid one child, every other vertex in $C$ precisely two children etc .. }

\begin{lemma}\label{lm:mdtclique}
Let $G$ be a \gatex graph,  $(N,t)$ be a labelled galled-tree explaining $G$ and $v$ be a vertex of $N$ that is not the root $\rho_C$ of any $C$ of $N$.
Moreover, let $L_u = L(N(u))$ for $u\in V(N)$.
Then it holds that 
\begin{enumerate}
\item If $t(v)=0$, then $\omega(G[L_v])=\max_{w \in \child_N(v)}\{\omega(G[L_w])\}$ and any maximum clique of $G[L_v]$ is entirely contained in $G[L_w]$ for some $w \in \child_N(v)$.
\item  If $t(v)=1$, then $\omega(G[L_v])=\sum_{w \in \child_N(v)} \omega(G[L_w])$ and any maximum clique of $G[L_v]$ is the join union $\join_{w \in \child_N(v)} K^w$
of maximum cliques $K^w$ in $G[L_w]$.
\end{enumerate}
\end{lemma}

\begin{proof}
Since no cycle $C$ of $N$ satisfies $\rho_C=v$, one easily verifies that, for all distinct $z', z \in \child_N(v)$,
it holds that $L_z \cap L_{z'}=\emptyset$ and that $\lca_N(x,x')=v$ for all $x \in L_z, x' \in L_{z'}$ (see also \cite[Lemma 2.1]{HS:22}).
Since $(N,t)$ explains $G$, it follows that $\{x,x'\}$ is an edge of $G$ if and
only if $t(v)=1$. In particular, the following holds:

\emph{Case $t_G(v)=0$:} In this case, $G[L_v]$ is the disjoint union of the graphs $G[L_w]$ with
$w\in \child_N(v)$. Hence, every maximum clique in $G[L_v]$ must be located entirely in one of the
subgraphs $G[L_w]$ of $G[L_v]$. Consequently, $\omega([G[L_v])=\max_{w \in
\child_N(v)}\{\omega(G[L_w])\}$ holds. 

\emph{Case $t_G(v)=1$:} Suppose that $K$ is a maximum clique in $G[L_v]$.
Since $t_G(v)=1$, $G[L_v]$ is the join union of the graphs $G[L_w]$ with $w\in \child_N(v)$. 
In particular, $K$ can be written as the join union of cliques $K^w$ in
$G[L_w]$, $w\in \child_N(v)$. Note that each of the cliques $K^w$ must be 
a maximum clique in $G[L_w]$ as otherwise we can replace $K^w$  by a larger clique
in $G[L_w]$ and obtain a clique $K'$ in $G[L_v]$ that is larger than $K$.
Consequently, $\omega(G[L_v])=\sum_{w \in \child_N(v)} \omega(G[L_w])$ holds.
\end{proof}

We next investigate the case of vertices $v$ of $\MDT_G$ with $t_G(v)=\mathit{prime}$.
To recall, 
we denote with $P^1(C),P^2(C)$ the sides of cycles $C\subseteq N$, 
i.e., the two directed paths $C$ with the same start-vertex $\rho_C$ and
end-vertex $\eta_C$, and whose vertices distinct from $\rho_C$ and $\eta_C$ are
pairwise distinct. Moreover
	 $G_1(M), G_2(M)\subseteq G[M]$ will denote the subgraphs induced by leaf-descendants of
	the vertices in $P^1(C_M)-\rho_{C_M}$ and $P^2(C_M)-\rho_{C_M}$, respectively.

In the upcoming proofs we may need to compute the join $H'\join H$ where $H$ is the empty graph. 
To avoid cumbersome case studies, we simple assume, in this case, that $H'\coloneqq H'\join H = H\join H'$. 
In other words, if $H$ is empty and we argue along $H'\join H$, then all arguments are 
applied to $H'$.

\begin{lemma}\label{lem:max-clique-in-eta}
	Let $G$ be a \gatex graph that is explained by the pvr-network $(N,t)$ and
	suppose that $G$ contains a prime module $M$. Put $L_{\eta}\coloneqq
	L(N(\eta_{C_M}))$ and let $H\in \{G[M], G_1(M), G_2(M)\}$.
	If $H$ contains a maximum clique $K$ with
	vertices in $L_\eta$, then $V(K)\cap L_\eta$ induces a maximum clique in
	$G[L_\eta]$ and $(V(K)\setminus L_\eta)\cupdot V(K')$
	induces a maximum clique in $H$ for every   maximum clique $K'$ in $G[L_\eta]$. 
\end{lemma}
\begin{proof}
	Let $G$ be a \gatex graph that is explained by the pvr-network $(N,t)$ and
	suppose that $G$ contains a prime module $M$. Put $L_{\eta}\coloneqq
	L(N(\eta_{C_M}))$, $\eta \coloneqq \eta_{C_M}$ and $p=\rho_{C_M}$. In the following, let 
	$H\in \{G[M], G_1(M), G_2(M)\}$. 

	Suppose that $H$ contains a maximum clique $K$ that contains vertices in
	$L_\eta$. Since $K$ is a clique in $H$, it must hold that $t(\lca_N(x,z)) =
	1$ for all $x \in V(K)\cap L_\eta$ and $z\in V(K)\setminus L_\eta\subseteq
	V(H)\setminus L_\eta$. By definition of pvr-networks, $L_\eta$ is a module
	of $G$ and, therefore, $t(\lca_N(x,z)) = 1$ with $x \in V(K)\cap L_\eta$ and
	$z\in V(K)\setminus L_\eta$ implies that $t(\lca_N(x',z)) = 1$ for all $x'
	\in L_\eta$. By construction, we have $V(K) = (V(K)\setminus L_\eta)\cupdot
	(V(K) \cap L_\eta)$. Assume, for contradiction, that $V(K)\cap L_\eta$ does
	not induce a maximum clique in $G[L_\eta]$. In this case, there is a clique
	$K'$ in $G[L_\eta]$ such that $|V(K')|>|V(K)\cap L_\eta|$. By the previous
	arguments, $t(\lca_N(x',z)) = 1$ for all $x' \in V(K')$ and $z\in
	V(K)\setminus L_\eta$. This together with $G[L_\eta]\subseteq H$
	  implies that $(V(K)\setminus L_\eta)\cupdot
	V(K')$ induces a complete graph in $H$. However, 
	 $|(V(K)\setminus
	L_\eta)\cupdot V(K')|> |(V(K)\setminus L_\eta)\cupdot (V(K) \cap L_\eta)| =
	|V(K)|$; a contradiction to $K$ being a maximum clique in $H$. 	
	Therefore, $V(K)\cap L_\eta$ induces a maximum clique in $G[L_\eta]$.
	
	Finally, let $K'$ be some maximum clique in $G[L_\eta]$ and thus, $|V(K)\cap L_\eta|=|V(K')|$. 
	As argued before, $t(\lca_N(x',z)) = 1$ for all $x' \in V(K')$ and $z\in
	V(K)\setminus L_\eta$ which implies that $(V(K)\setminus L_\eta)\cupdot
	V(K')$ induces a complete graph $K''$ in $H$ of size 
	$|V(K'')| = |V(K)\setminus L_\eta| + |V(K')| = |V(K)\setminus L_\eta| + |V(K)\cap L_\eta| = |V(K)|$.
	Hence, $K''$ is a maximum clique in $H$. 
\end{proof}

\begin{lemma}\label{lem:Leta2}
	Let $G$ be a \gatex graph that is explained by the pvr-network $(N,t)$ and
	suppose that $G$ contains a prime module $M$ such that $t(\rho_{C_M})=1$. If
	$G[M]$ contains a maximum clique that contains vertices in
	$L(N(\eta_{C_M}))$, then $G_1(M)$ and $G_2(M)$ have both a maximum clique
	that contains vertices in $L(N(\eta_{C_M}))$.
\end{lemma}
\begin{proof}
	Let $G$ be a \gatex graph that is explained by the pvr-network $(N,t)$ and
	suppose that $G$ contains a prime module $M$ such that $t(\rho_{C_M})=1$.
	Furthermore, put $G_1\coloneqq G_1(M)$, $G_2\coloneqq G_2(M)$,
	$L_{\eta}\coloneqq L(N(\eta_{C_M}))$ and $G_\eta\coloneqq G[L_\eta]$. For a
	subgraph $H\subseteq G$ we define $|H|\coloneqq |V(H)|$. Let $K$ be a
	maximum clique in $G[M]$ that contains vertices in $L_\eta$ and put $K^1
	\coloneqq (G_1 - G_\eta) \cap K$, $K^2 \coloneqq (G_2 - G_\eta) \cap K$ and
	$K^\eta \coloneqq K \cap G_\eta$. Thus, $V(K) = V(K^1)\cupdot
	V(K^\eta)\cupdot V(K^2)$. 
	
	Assume, for contradiction, that every maximum clique in $G_1$ does not
	contain vertices in $L_\eta$. Let $K'$ be a maximum clique in $G_1$. Since
	$V(K^1)\cupdot V(K^\eta)\subseteq V(G_1)$ and $V(K^1)\cupdot V(K^\eta)$
	induce a complete graph with vertices in $L_\eta$, we can conclude that
	$|V(K^1)\cupdot V(K^\eta)| = |K^1| +|K^\eta|<|K'|$. Note that
	$\lca_N(x,y)=\rho_C$ has label $1$ for all $x\in V(K')$ and $y\in V(K^2)$
	and thus, $K'' \coloneqq K'\join K^2$ forms a complete graph in $G[M]$ and
	thus, $|K'| +|K^2| = |K''| \leq |K|$. This together with $|K^1|
	+|K^\eta|<|K'|$ yields the following contradiction: \[ |K'|+|K^2| =
	|K''|\leq |K| = |K^1|+|K^\eta|+|K^2|< |K'|+|K^2|.\] Hence, $G_1$ must
	contain a maximum clique with vertices in $L_\eta$. By similar arguments,
	$G_2$ must contain a maximum clique with vertices in $L_\eta$. 
\end{proof}

%Lemma \ref{lem:Leta1} and \ref{lem:Leta2} imply
%\begin{proposition}\label{prop:Leta}
%	Let $G$ be a \gatex graph that is explained by the pvr-network $(N,t)$
%	and suppose that $G$ contains a prime module $M$ such that $t(\rho_{C_M})=1$. 
%	Then, the following two statements are equivalent:
%	\begin{enumerate}
%		\item $G_1(M)$ and $G_2(M)$ contain a maximum clique  with vertices in $L(N(\eta_C))$.
%		\item $G[M]$ contains a maximum clique with vertices in $L(N(\eta_C))$.
%	\end{enumerate}
%\end{proposition}

\begin{lemma}\label{lem:not-Leta2}
	Let $G$ be a \gatex graph that is explained by the pvr-network $(N,t)$ and
	suppose that $G$ contains a prime module $M$ such that $t(\rho_{C_M})=1$. Put
	$L_\eta \coloneqq L(N(\eta_C))$ and $G_\eta = G[L_\eta]$. Furthermore, suppose
	that $G_1(M)$, resp., $G_2(M)$ have a maximum clique $K'$, resp., $K''$ with
	vertices in $L_\eta$ and such that $V(K')\cap L_\eta = V(K'')\cap L_\eta$. If
	$V(K')\cup V(K'')$ does not induce a maximum clique in $G[M]$, then none of
	the maximum cliques in $G[M]$ can have vertices in $L_\eta$. 
\end{lemma}
\begin{proof}
	Let $G$ be a \gatex graph that is explained by the pvr-network $(N,t)$ and
	suppose that $G$ contains a prime module $M$ such that $t(\rho_{C_M})=1$.
	Furthermore, put $G_1\coloneqq G_1(M)$, $G_2\coloneqq G_2(M)$,
	$L_{\eta}\coloneqq L(N(\eta_{C_M}))$ and $G_\eta\coloneqq G[L_\eta]$. For a
	subgraph $H\subseteq G$ we define $|H|\coloneqq |V(H)|$. Suppose that $G_1$,
	resp., $G_2$ contains a maximum clique $K'$, resp., $K''$ that contains
	vertices in $L_{\eta}$. 
	Assume first that $V(K')\cap L_{\eta} \neq V(K'') \cap L_{\eta}$. 
		By Lemma \ref{lem:max-clique-in-eta}, $V(K')\cap L_\eta$ 
		and $V(K'')\cap L_\eta$ induce a maximum clique in 
		$G[L_\eta]$ and $(V(K')\setminus L_\eta)\cupdot (V(K'')\cap L_\eta)$	
		induces a maximum clique $K'''$ in $G_1$ with vertices in $L_\eta$. 
		In particular, $V(K''')\cap L_{\eta} = V(K'') \cap L_{\eta}$
		is satisfied. 
		
		Hence, we can assume in the following w.l.o.g.\
		that $V(K')\cap L_{\eta} = V(K'') \cap L_{\eta}$.
		By Lemma \ref{lem:max-clique-in-eta},  $V(K')\cap L_{\eta} = V(K'') \cap L_{\eta}$ induces a maximum
	clique $K^\eta$ in $G[L_\eta]$. 
	
	We show first that $V(K')\cup V(K'')$ induces a complete graph in $G[M]$. Let
	$K^1$, resp., $K^2$ be the complete subgraph of $K'$, resp., $K''$ that is
	induced by $V(K')\setminus L_{\eta}$, resp., $V(K'')\setminus L_{\eta}$. Since
	$t(p)=1$, all vertices in $V(K^1)$ are adjacent to all vertices in $V(K^2)$
	and thus, the subgraph induced by $V(K^1) \cupdot V(K^2)$ coincides with
	$K^1\join K^2$. Since $K^\eta$ is a complete graph, we have $K' = K^1\join
	K^\eta$ and $K'' = K^2\join K^\eta$, The latter two arguments imply that $K'''
	\coloneqq K^1\join K^\eta\join K^2$ is a complete graph in $G[M]$ that is
	induced by $V(K')\cup V(K'')$.

	Suppose now that  $K'''$ is not a maximum clique in $G[M]$. Let $\hat K$
	be a maximum clique in $G[M]$ and thus, $|\hat K|>|K'''|$. Assume, for
	contradiction, that $\hat K$ contains vertices in $L_\eta$. Thus, we can apply
	Lemma \ref{lem:max-clique-in-eta} and assume w.l.o.g.\ that $V(\hat K)\cap
	L_\eta = V(K^\eta)$ induces the maximum clique $K^\eta$ in $G[L_\eta]$. Let
	$\hat K^i$ be the complete subgraph of $\hat K$ induced by the vertices
	$(V(G_i)\cap V(\hat K)) \setminus L_{\eta}$, $i\in \{1,2\}$. By construction,
	$V(K''') = V(K^1)\cupdot V(K^\eta)\cupdot V(K^2)$ and, $V(\hat K) = V(\hat
	K^1)\cupdot V(K^\eta)\cupdot V(\hat K^2)$. If $|\hat K^1|\leq |K^1|$ and
	$|\hat K^2|\leq |K^2|$, then $|\hat K| = |\hat K^1| + |K^\eta| + |\hat
	K^2|\leq |K^1| + |K^\eta| + |K^2| = |K'''|$, which is impossible as, by
	assumption, $|\hat K|>|K'''|$. Thus, $|\hat K^1|>|K^1|$ or $|\hat K^2|>|K^2|$
	must hold. W.l.o.g.\ we may assume that $|\hat K^1|>|K^1|$. But then, $|V(\hat
	K^1)\cupdot V(K^\eta)|> |V(K^1)\cupdot V(K^\eta)|$ which together with the
	fact that $V(\hat K^1)\cupdot V(K^\eta)\subseteq V(\hat K)$ induce a complete
	graph in $G_1$ implies that $V(K^1)\cupdot V(K^\eta) = V(K')$ cannot induce a
	maximum clique in $G_1$; a contradiction. Thus, $\hat K$ cannot contain
	vertices in $L_\eta$.
\end{proof}

\begin{proposition}\label{prop:clique-Number-M-series}
	Let $G$ be a \gatex graph that is explained by the pvr-network $(N,t)$ and
	suppose that $G$ contains a prime module $M$ where $t(\rho_{C_M})=1$. Put
	$L_\eta = L(N(\eta_{C_M}))$, $G_1=G_1(M)$, $G_2=G_2(M)$ and $G_\eta =
	G[L_\eta]$. Then, \[\omega(G[M]) =
	\max\{\omega(G_1)+\omega(G_2)-\omega(G_\eta),
	\omega(G_1-G_\eta)+\omega(G_2-G_\eta)\}.\]

	\noindent
	In particular, the following statements hold for $\alpha\coloneqq
	\omega(G_1)+\omega(G_2)-\omega(G_\eta)$ and $\beta\coloneqq
	\omega(G_1-G_\eta)+\omega(G_2-G_\eta)$:
	\begin{enumerate}
		\item If $\alpha\leq \beta$, then  $K = K_1\join K_2$ is a maximum clique in $G[M]$
					for	every maximum clique $K^1$ in $G_1-G_\eta$ and $K^2$ in   $G_2-G_\eta$.
		\item If $\alpha>\beta$, then every maximum clique in $G[M]$ contains vertices in $L_\eta$ and
						$V(K')\cup V(K'')$
					induces a maximum clique in $G[M]$ for 
					every maximum clique $K'$ in $G_1$ and $K''$ in $G_2$ that satisfies
					$V(K')\cap L_\eta= V(K'')\cap L_\eta\neq \emptyset$.
	\end{enumerate}
\end{proposition}
\begin{proof}
	Let $G$ be a \gatex graph that is explained by the pvr-network $(N,t)$ and
	suppose that $G$ contains a prime module $M$ where $t(\rho_{C_M})=1$. Put
	$L_\eta = L(N(\eta_{C_M}))$, $G_1=G_1(M)$, $G_2=G_2(M)$ and $G_\eta =
	G[L_\eta]$. Let $K$ be a maximum clique in $G[M]$. For a subgraph $H\subseteq
	G$ we define $|H|\coloneqq |V(H)|$.
	
	We start with showing that $\omega(G[M]) =
	\max\{\omega(G_1)+\omega(G_2)-\omega(G_\eta),
	\omega(G_1-G_\eta)+\omega(G_2-G_\eta)\}$. 
	Since  $\lca(x,y)=\rho_{C_M}$ and
	$t(\rho_{C_M})=1$ for all $x\in V(G_1-G_\eta)$ and $y\in V(G_2-G_\eta)$, 
	every complete subgraph $K^1$ in $G_1-G_\eta$ and $K^2$ in $G_2-G_\eta$ yields
	a complete subgraph $K^1\join K^2$ in $G[M]$.
	Consider two maximum cliques $K^1$ in $G_1-G_\eta$ and $K^2$ in $G_2-G_\eta$. Hence,
	$\omega(G_1-G_\eta) = |K^1|$ and $\omega(G_2-G_\eta) = |K^2|$. Moreover, $\tilde K
	\coloneqq K^1\join K^2$ forms a complete graph in $G[M\setminus
	L_\eta]\subseteq G[M]$. Therefore, $|K|\geq |\tilde K| = |K_1|+|K_2|$ and,
	thus, $\omega(G[M])\geq \omega(G_1-G_\eta)+\omega(G_2-G_\eta)$. Hence, if
	$\omega(G[M]) = \omega(G_1-G_\eta)+\omega(G_2-G_\eta)$ we are done. Assume that
	$\omega(G[M]) > \omega(G_1-G_\eta)+\omega(G_2-G_\eta)$. 
	Let $\tilde K$ be a maximum clique in $G[M\setminus L_\eta]$
	and assume, for contradiction, that $\tilde K$ is  a maximum clique in $G[M]$.
	By similar arguments as before, we can write $\tilde K = \tilde K^1\join \tilde K^2$
	where $\tilde K^1 = \tilde K\cap (G_1-G_\eta)$ and $\tilde K^2 = \tilde K\cap (G_2-G_\eta)$. 
	In particular, $\tilde K^1$ must be a maximum clique in $G_1-G_\eta$
	since, otherwise, there is a larger clique $\tilde K'$ in $G_1-G_\eta$ 
	and, thus, $\tilde K'\join  K_2$ would be larger
	than $\tilde K = \tilde K^1\join \tilde K^2$; a contradiction. 
	Similarily, $\tilde K^2$ must be a maximum clique in $G_2-G_\eta$. 
	Hence, $\omega(G[M]) = |\tilde K| = |\tilde K^1| + |\tilde K^2| = \omega(G_1-G_\eta)+\omega(G_2-G_\eta)<\omega(G[M])$;
	a contradiction. Hence, 
	none of the complete graphs in $G[M\setminus L_\eta]$ are maximum cliques in
	$G[M]$. This and the fact that $M = V(G_1-G_\eta)\cupdot V(G_2-G_\eta) \cupdot
	V(G_\eta)$ implies that maximum cliques in $G[M]$ and, therefore, $K$ must
	contain vertices in $L_\eta$. By Lemma \ref{lem:Leta2}, $G_1$ and $G_2$ have
	both a maximum clique with vertices in $L_\eta$. Let $K'$, resp., $K''$ be a
	maximum clique in $G_1$, resp., $G_2$. Let $K^1$, resp., $K^2$ be the complete
	subgraph of $K'$, resp., $K''$ that is induced by $V(K')\setminus L_{\eta}$,
	resp., $V(K'')\setminus L_{\eta}$. By Lemma \ref{lem:max-clique-in-eta}, we
	can assume w.l.o.g.\ that $V(K')\cap L_{\eta} = V(K'') \cap L_{\eta}$ induce a
	maximum clique $K^\eta$ in $G[L_\eta]$. Hence, $K''' = K^1\join K^\eta \join
	K^2$ is a complete subgraph of $G[M]$. In particular, $K'''$ is induced by
	$V(K')\cup V(K'')$. Thus, 
	if $K''' $ is not a maximum clique in
	$G[M]$, then  Lemma \ref{lem:not-Leta2} implies that none of
	the maximum cliques in $G[M]$ can have vertices in $L_\eta$; a contradiction.
	Hence, $K'''$ is a maximum clique in $G[M]$. Since $V(K''') =
	V(K')\cup V(K'')$ and $V(K')\cap V(K'')=V(K^\eta)$, we obtain $\omega(G) =
	|K'''| = |K'| + |K''| - |K^\eta| = \omega(G_1)+\omega(G_2)-\omega(G_\eta)$. In
	summary, $\omega(G[M]) = \max\{\omega(G_1)+\omega(G_2)-\omega(G_\eta),
	\omega(G_1-G_\eta)+\omega(G_2-G_\eta)\}$.	
	
	We now verify the Conditions (1) and (2) in the second statement. Consider
	first Condition (1) and assume that $\alpha \coloneqq
	\omega(G_1)+\omega(G_2)-\omega(G_\eta)\leq
	\omega(G_1-G_\eta)+\omega(G_2-G_\eta) \eqqcolon \beta$. By the previous
	arguments, $\omega(G[M]) = \beta$. Let $K^1$ be a maximum clique in
	$G_1(M)-G_\eta$ and $K^2$ be a maximum clique in $G_2(M)-G_\eta$ and thus,
	$\omega(G_1-G_\eta) = |K^1|$ and $\omega(G_2-G_\eta) = |K^2|$. By the
	arguments above, $K=K^1\join K^2$ is a complete subgraph in $G[M]$. In
	particular, $|K| =
	|K^1|+|K^2|=\omega(G_1-G_\eta)+\omega(G_2-G_\eta)=\beta=\omega(G[M])$.
	Consequently, $K$ is a maximum clique in $G[M]$. Thus, Condition (1) is
	satisfied. 	
	
	For Condition (2), suppose that  $\alpha > \beta$ and, therefore,
	$\omega(G[M]) = \alpha$. As argued above,
	$\omega(G[M])>\omega(G_1-G_\eta)+\omega(G_2-G_\eta)$ implies that every
	maximum clique in $G[M]$ must contain vertices in $L_\eta$. This and Lemma
	\ref{lem:Leta2} implies that $G_1$ and $G_2$ have both a maximum clique that
	contains vertices in $L_\eta$. Let $K'$ be an arbitrary maximum clique in
	$G_1$ and $K''$ be an arbitrary maximum clique in $G_2$ such that $V(K')\cap
	L_\eta= V(K'')\cap L_\eta\neq \emptyset$ holds. By Lemma
	\ref{lem:max-clique-in-eta}, such cliques $K'$ and $K''$ exist. Since every
	maximum clique in $G[M]$ must contain vertices in $L_\eta$, contraposition of
	Lemma \ref{lem:not-Leta2} implies that $V(K')\cup V(K'')$ induce a maximum
	clique in $G[M]$. Thus, Condition (2) is satisfied. 	
\end{proof}

\begin{proposition}\label{prop:clique-Number-M-parallel}	
	Let $G$ be a \gatex graph that is explained by the pvr-network $(N,t)$
	and suppose that $G$ contains a prime module $M$. If $t(\rho_{C_M})=0$, then
	$\omega(G[M]) = \max\{\omega(G_1(M)),\omega(G_2(M))\}$ of \gatex graphs 
\end{proposition}
\begin{proof}
	Let $G$ be a \gatex graph that is explained by the pvr-network $(N,t)$
	and suppose that $G$ contains a prime module $M$ such that 
	that $t(\rho_{C_M})=0$. Furthermore, put $L_{\eta}\coloneqq L(N(\eta_{C_M}))$ and $G_\eta\coloneqq G[L_\eta]$. 
	Let $K$ be a maximum clique in $G[M]$. Note first that $K$ cannot contain 
	vertices $x$ and $y$ such that $x\in V(G_1(M)-G_\eta)$ and $y\in V(G_2(M)-G_\eta)$ 
	since, in this case,  $\lca(x,y)=\rho_{C_M}$ and $t(\rho_{C_M})=0$ imply that $\{x,y\}\notin E(G[M])$. 
	Hence, $K$ must be entirely contained in either $G_1(M)$ or $G_2(M)$. 
	Moreover, any maximum clique in $G_1(M)$ and $G_2(M)$ provide a complete
	subgraph of $G[M]$. Taken the latter two arguments together, 
	$\omega(G[M]) = \max\{\omega(G_1(M)),\omega(G_2(M))\}$.
\end{proof}

\begin{remark}
For the sake of simplicity, we often 
put $\omega(w)\coloneqq \omega(G[L_w])$ for 
the size of a maximum clique in the subgraph of $G$ induced by $L_w = L(N(w))\subseteq V(G)$.
where $(N,t)$ is a galled-tree that explains $G$. 
\end{remark}

\begin{algorithm}[t]
  \small %\footnotesize % \scriptsize
  \caption{\texttt{Computation of a maximum clique and $\omega(G)$ of \gatex graphs $G$}  }
  \label{alg:clique}
  \begin{algorithmic}[1]
    \Require  A \gatex graph $G=(V,E)$ %\Comment{Highlighted blue-colored text can be ignored if only $\omega(G)$ is of interest}
    \Ensure   maximum clique $K$ in $G$  and its size $\omega(G)$ %\ \medskip\medskip

    \State Compute $(\MDT_G,t_G)$ and pvr-network $(N,t)$ of $G$ \label{l:MDT-pvr}
    \State 		put $\omega(v)\coloneqq 1$ for all leaves $v$ in $L(N)=V$   \label{l:omega-leaves}
    
    \ForAll{$v\in V(\MDT_G)\setminus{L(\MDT_G)}$ in postorder} \label{l:forV}
    
    	\If{$t_G(v)=0$}
    	 	\State Put $\omega(v) \coloneqq \max_{w\in \child_{\MDT_G}(v)} \{\omega(w)\}$ \label{l:v-parallel}
    	 	\State Mark $w$ as \texttt{active} for precisely one $w\in \argmax_{z\in \child_{\MDT_G}(v)} \{\omega(z)\}$ \label{l:active-0}

    	\ElsIf{$t_G(v)=1$}	
    		\State Put $\omega(v) \coloneqq \sum_{w\in \child_{\MDT_G}(v)} \omega(w)$ \label{l:v-series}
    	 	\State Mark all  $w\in \child_{\MDT_G}(v)$ as \texttt{active}  \label{l:active-1}

    	\Else  \Comment{$t_G(v)=\mathrm{prime}$ } \label{l:v-prime}
    	 	\State Let $C$ be the unique cycle in $N$ with root $\rho_C=v$  \label{l:cycle2}
    	 	
    	 			\LComment{Although $\rho_C=v$,  we distinguish between them to make it clearer if we are working in  $\MDT_G$ or $N$}			%\Comment{$L(N(\rho_C)) = L(\MDT_G(v))$}

    	 	\State Let $\eta$ be the unique hybrid in $C$ and 
    	 			   $u$ be the unique child of $\eta$ in $N$ \label{l:eta}
    	 	\State Put $\omega(\eta)\coloneqq \omega(u)$ and $\omegaExclHyb(\eta) \coloneqq 0$  \label{l:omega-eta} \	\smallskip \smallskip 		   

			\LComment{Init  $\omega(w)$ and $\omegaExclHyb(w)$ for the vertices $w\neq \rho_c,\eta$ along the sides of $C$ bottom-up}
			\State Let $P^1$ and $P^2$ be the two sides of $C$ \label{l:sides}
			\ForAll{$w\in V(P^i)\setminus\{\rho_C, \eta\}$ in postorder, $i\in \{1,2\}$}  \label{l:init-innerC-start}
				
					\State Put $u'\coloneqq \child_N(w)\cap V(C)$ and $u''\coloneqq \child_N(w)\setminus V(C)$  \Comment{Note, $\child_N(w)=\{u',u''\}$ for $w\neq \eta$} \label{l:wu'u''}		
					
					\If{ $t(w) = 0$ } 
						Put $\omega(w) \coloneqq \max\{\omega(u'),\omega(u'')\}$ 
						and $\omegaExclHyb(w) \coloneqq \max\{\omegaExclHyb(u'),\omega(u'')\}$  \label{l:if-tw-0}					
					\Else	\
										Put $\omega(w) \coloneqq \omega(u') + \omega(u'')$
						and $\omegaExclHyb(w) \coloneqq \omegaExclHyb(u') + \omega(u'')$ \label{l:if-tw-1}					 \Comment{$t(w) = 1 $} 
					\EndIf
			\EndFor \label{l:init-innerC-end} \ \smallskip\smallskip

			\LComment{Init $\omega(v)$. Note, $\rho_C$ corresponds to $v$ in $\MDT_G$}
			\State Let $u'$ and $u''$ be the two children of $\rho_C$	 \label{l:rho1} 
			\If{$t(\rho_C) = 0$}  \label{l:if1}
				\State	 Put $\omega(v) \coloneqq \max\{\omega(u'), \omega(u'')\}$ \label{l:parallel}
				\State   Choose one $w\in\argmax\{\omega(u'), \omega(u'')\}$ \label{l:pick-w}
				\State   Mark $w$ as \texttt{active} \label{l:activate-picked}
				\State   Let $P\in \{P^1,P^2\}$ be such that $w\in V(P)$. \label{l:pick-P}
				\State  \textsc{Activate}($(N,t)$,$\{P\}$, $\omega$, $\omegaExclHyb$, 0, 0, 0) \label{l:activate-0}
				
			\Else  \Comment{$t(\rho_C) = 1$} \label{l:else1}
				\State Put $\alpha\coloneqq \omega(u')+\omega(u'')-\omega(\eta)$ and $\beta\coloneqq \omegaExclHyb(u') + \omegaExclHyb(u'')$ \label{l:series-1}
				\State Put $\omega(v) \coloneqq \max\{\alpha, \beta\}$ \label{l:series-2}
				\State		\textsc{Activate}($(N,t)$,$\{P^1,P^2\}$, $\omega$, $\omegaExclHyb$, $\alpha$, $\beta$, 1) \label{l:activate-1}
			\EndIf	\label{l:rho2}

    	\EndIf \label{l:end-v-prime}
	    		
    \EndFor \label{l:forV-end}

	\vspace{-0.25in}    
    \State $\Omega \coloneqq$ set of all leaves $x\in L(N)$ for which there is a path $P$ from $\rho_N$ to $x$  where all vertices $v\neq \rho_N$ in $P$ are \texttt{active}  \label{l:omega}
    \State \Return $G[\Omega]$ and $\omega(\rho_{N})$
  \end{algorithmic}
\end{algorithm}

\begin{algorithm}[htbp]
  \small %\footnotesize % \scriptsize
  \caption*{\textbf{Procedure} \textsc{Activate}$\left((N,t), \mathcal{P}, \omega, \omegaExclHyb, \alpha, \beta, \textit{label}\_\rho_C \right)$	}
  \label{alg:activate}
  \begin{algorithmic}[1]
		\If{$\textit{label}\_\rho_C=0$} 																																				\label{l:active:0}
			\For{$w'\in V(P) \setminus\{\rho_C\}$ in postorder where $P\in \mathcal{P}$} 																							\label{l:active:0:P}
				\State Put $u'\coloneqq \child_N(w')\cap V(C)$ and $u''\coloneqq \child_N(w')\setminus V(C)$ 					\label{l:active:0:u'u''} 
		    	\If{$t(w')=0$}  Mark precisely one $u\in \argmax\{\omega(u'),\omega(u'')\}$ as \texttt{active} 	\label{l:active:0:0} 
		    	 \Else \ Mark all $u\in \child_N(w')$ as \texttt{active} 																											\label{l:active:0:1} 
		    	\EndIf	
			\EndFor 
		\Else \  Mark both children of $\rho_C$ as \texttt{active} 																							\label{l:active:1}
		\If{$\alpha\leq \beta$} 																																								\label{l:active:1:aLEQb}
				\State Let $w_1$ and $w_2$ be the unique parents of  $\eta$ 																					\label{l:active:1:eta}
			\State Mark $u\in \child(w_i)\setminus \{\eta\}$ as \texttt{active} for $i\in \{1,2\}$ 								\label{l:active:1:parent-eta} 
			\For{all $P\in \mathcal{P}$ and $w'\in V(P)\setminus \{\rho_C,\eta,w_1,w_2\}$ in postorder}  					\label{l:active:1:P}
				\State Put $u'\coloneqq \child_N(w')\cap V(C)$ and $u''\coloneqq \child_N(w')\setminus V(C)$ 					\label{l:active:1:u'u''} 
				\If{ $t(w') = 0$ } 																																									\label{l:active:1:0} 
					   Mark exactly one $u\in\argmax\{\omegaExclHyb(u'), \omega(u'')\}$ as \texttt{active} \label{l:pick-u}	
%					\If{$\omegaExclHyb(u')\geq \omega(u'')$} Mark $u'$ as \texttt{active} 														\label{l:active:1:u'-active} 
%					\Else\ Mark $u''$ as \texttt{active} 																															\label{l:active:1:u''-active} 
%					\EndIf				
				\Else \ 
%					\State\TODO{NEEDED?}	Put $\omega(w) \coloneqq \omega(u') + \omega(u'')$ and 
%								$\omegaExclHyb(w) \coloneqq \omegaExclHyb(u') + \omega(u'')$ 													\label{l:active:1:u'u''2} 
					  Mark both children $u',u''$ of $w'$ as \texttt{active} 																							\label{l:active:1:active} 
					 \Comment{$t(w) = 1 $} 																																			\label{l:active:1:1} 
				\EndIf
			\EndFor
	 	\Else \ \Comment{$\alpha> \beta$}																																				\label{l:active:1:a>b}
	 		\State Mark $\eta_C$ and its child as \texttt{active} \label{l:active:1:eta}
			\For{all $P\in \mathcal{P}$ and $w\in V(P)\setminus\{\rho_C,\eta_C\}$ in postorder}													\label{l:active:1:P2}
				\State Mark $w$ as  \texttt{active}																																	\label{l:active:1:w-active} 
				\If{ $t(w) = 1$ } mark also the child of $w$ not in $P$ as  \texttt{active}													\label{l:active:1:w=1} 
				\EndIf	
			\EndFor
		\EndIf
	  \EndIf	
  \end{algorithmic}
\end{algorithm}

\begin{figure}[t]
	\begin{center}
			\includegraphics[width=.6\textwidth]{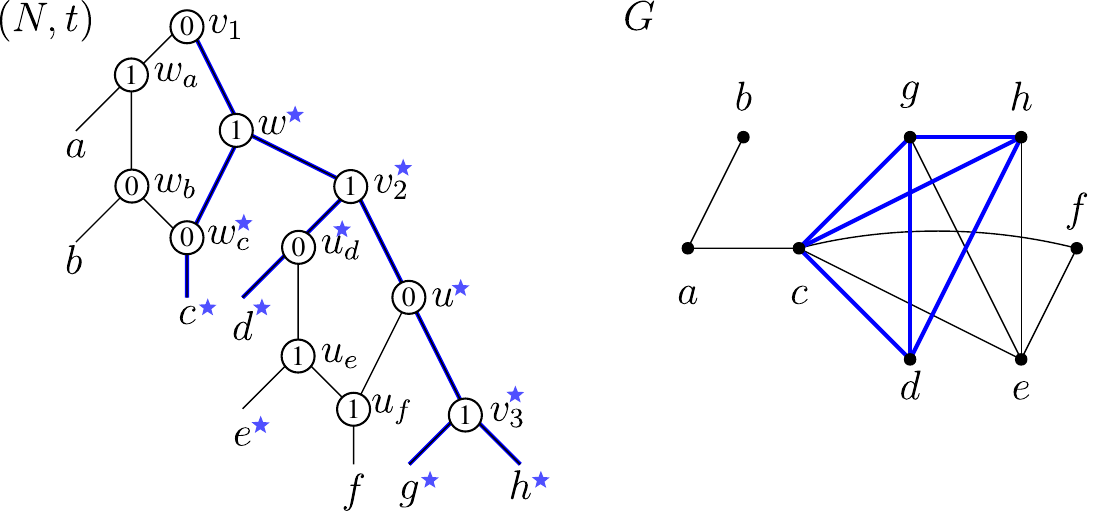}
	\end{center}
	\caption{Left a  galled-tree $(N,t)$ that explains the \gatex graph $G$ on the right. 
					Algorithm~\ref{alg:clique} returns the induced subgraph $G[g,h,d,c]$ (highlighted in blue) which is  a maximum clique of $G$. 
					All vertices marked as \texttt{active} are highlighted with $\star$. 
					Paths $P$ from $\rho_N$ to leaves $x$ in $N$ where all vertices $v\neq \rho_N$ in $P$ are \texttt{active} are highlighted in blue;
					see explanations in Example \ref{exmpl:alg} for further details.}
\label{fig:example-clique}
\end{figure}

As we shall see, Algorithm~\ref{alg:clique} can be used to compute maximum cliques in \gatex graphs in linear-time. 
Before studying Algorithm~\ref{alg:clique} in detail, we illustrate this
algorithm on the examples as shown in Figure~\ref{fig:example-clique}.

\begin{example}\label{exmpl:alg}  
We exemplify here the main steps of Algorithm~\ref{alg:clique} using the \gatex graph $G$ as shown
in Fig.~\ref{fig:example-clique}. We first compute the modular decomposition tree $(\MDT_G,t_G)$ 
(as shown in Fig.~\ref{fig:example}) and
the shown pvr-network $(N,t)$ that explains $G$ (Line \ref{l:MDT-pvr}). For all leaves $v$ of $\MDT_G$ (and
thus, of $N$), we have $L_v=\{v\}$ and, thus, the size of a maximum clique in $G[v]$ is one and
we put $\omega(v)\coloneqq 1$ (Line~\ref{l:omega-leaves}). We then traverse the vertices $\MDT_G$
that are not leaves in postorder and thus obtain the order $v_3,v_2,v_1$ in which the vertices are
visited (Line~\ref{l:forV}). Note that postorder-traversal ensures that all children of a given
vertex $v$ in $\MDT_G$ are visited before this vertex $v$ is processed. 
In what follows, we denote with $\omegaExclHyb(v)$ the size of a maximum clique
in $G[L_v\setminus L_{\eta_C}]$ given that $v$ is part of a cycle $C$ with hybrid $\eta_C$. 
%In other words, $\omegaExclHyb(v) = \omega(G[L(N(v))\setminus L(N(\eta_C)])$ is a maximum clique

Consider now the processed vertex $v_3$. Since $t(v_3)=1$, we define
$\omega(v_3)=\omega(g)+\omega(h)=2$ (Line~\ref{l:v-series}). The latter is in accordance with
Lemma \ref{lm:mdtclique} and refers to the fact that the maximum clique in $G[L_{v_3}]$ is precisely
the edge connecting $g$ and $h$, i.e, the join union of two single vertex graphs. In addition, we
mark both $g$ and $h$ as \texttt{active} (Line~\ref{l:active-1}). 

%The next processed vertex $v_2$ is a prime vertex in $\MDT_G$\mh{, i.e., we are now in
%Line~\ref{l:v-prime}.} We then consider the cycle $C$ \mh{in $N$} with root $\rho_C=v_2$ \mh{that is
%induced by $v_2, u,u_d,u_e,u_f$} (Line~\ref{l:cycle2}). 

We have $\eta_C=u_f$, so we put $\omega(u_f)=\omega(f)=1$ and $\omegaExclHyb(u_f)=0$
(Line~\ref{l:omega-eta}). The latter refers to the fact that, in this example, any maximum clique
in $G[L_\eta]$ is of size one, while any clique in $G[L_\eta\setminus L_\eta]$ is of size $0$.
The two sides of $C$ are $P^1=\{v_2,u,u_f\}$ and
$P^2=\{v_2,u_d,u_e,u_f\}$. We first consider the unique vertex $u$ of $P^1 \setminus
\{\rho_C,\eta_C\}$. We have $t(u)=1$, so we put $\omega(u)=\omega(u_f)+\omega(v_3)=3$, and
$\omegaExclHyb(u)=\omegaExclHyb(u_f)+\omega(v_3)=2$ (Line~\ref{l:if-tw-1}). Next, we consider the
vertices of $P^2 \setminus \{\rho_C,\eta_C\}=\{u_e,u_d\}$ in postorder, that is, $u_e$ first and
$u_d$ second. We have $t(u_e)=1$, so we put $\omega(u_e)=\omega(u_f)+\omega(e)=2$, and
$\omegaExclHyb(u_e)=\omegaExclHyb(u_f)+\omega(e)=1$ (Line~\ref{l:if-tw-1}). Since $t(u_d)=0$, we put
$\omega(u_d)=\max\{\omega(u_e),\omega(d)\}=2$, and
$\omegaExclHyb(u_e)=\max\{\omegaExclHyb(u_e),\omega(d)\}=1$ (Line~\ref{l:if-tw-0}). Afterwards, we
consider the vertex $\rho_C=v_2$. We have $t(v_2)=1$, so we go to Line~\ref{l:else1}. We first
define $\alpha=\omega(u_d)+\omega(u)-\omega(u_f)=3$, $\beta=\omegaExclHyb(u_d)+\omegaExclHyb(u)=3$
(Line~\ref{l:series-1}), and $\omega(v_2)=\max\{\alpha,\beta\}=3$ (Line~\ref{l:series-2}). 
The latter is in accordance with Prop.\ \ref{prop:clique-Number-M-series}.
In Line~\ref{l:activate-1}, we then call the procedure \textsc{Activate}$\left((N,t), \{P^1,P^2\}, \omega, \omegaExclHyb, 3, 3, 1
\right)$. This precedure is used to ``activate'' the right vertices in such a way that, after
termination of Algorithm~\ref{alg:clique}, the set $\Omega$ consisting of all leaves $x\in L(N)$ 
for which there is a path $P$ from $\rho_N$ to $x$ in $N$ with all vertices $v\neq \rho_N$ in $P$ marked as
\texttt{active} determines the vertex set of a maximum clique in $G$.

We are now in the procedure \textsc{Activate}. Since $label\_{\rho_C}=1$, we 
mark both children $u$ and $u_d$ of $\rho_C = v_2$ as \texttt{active} (Line~\ref{l:active-1}). Since $\alpha\leq \beta $,
we continue with Line~\ref{l:active:1:aLEQb}. In Line~\ref{l:active:1:parent-eta}, we mark $v_3$ and
$e$ as \texttt{active}. In the for-loop at Line~\ref{l:active:1:P}, we consider the 
two sides $P^1,P^2\in \mathcal{P}$ of $C$. In particular,
we have $V(P^1) \setminus \{\rho_C,\eta_C,u,u_e\}=\emptyset$ and $V(P^2)
\setminus \{\rho_C,\eta_C,u,u_e\}=\{u_d\}$. Thus,  we only have to consider, in this run of the for-loop,
 the vertex $w' = u_d$. In this case, $u'=u_e$ and $u''=d$.
Since $t(u_d)=0$, $\omegaExclHyb(u_e)=1$, and $\omega(d)=1$,
we choose one of $u_e$ or $d$ to be marked mark as \texttt{active} (Line~\ref{l:pick-u})
In this example, we decide to mark $d$ \texttt{active}. After this, 
we exit the procedure \textsc{Activate}.

We are now back in Algorithm~\ref{alg:clique} proceed with
the prime vertex $v_1$ and consider the cycle $C$ with root $\rho_C=v_1$
(Line~\ref{l:cycle2}). We have $\eta_C=w_c$, so we put $\omega(w_c)=\omega(c)=1$ and
$\omegaExclHyb(w_c)=0$ (Line~\ref{l:omega-eta}). The two sides of $C$ are $P^1=\{v_1,w,w_c\}$ and
$P^2=\{v_1,w_a,w_b,w_c\}$. We first consider the unique vertex $w$ of $P^1 \setminus
\{\rho_C,\eta_C\}$. We have $t(w)=1$, so we put $\omega(w)=\omega(w_c)+\omega(v_2)=4$, and
$\omegaExclHyb(w)=\omegaExclHyb(w_c)+\omega(v_2)=3$ (Line~\ref{l:if-tw-1}). Next, we consider the
vertices of $P^2 \setminus \{\rho_C,\eta_C\}=\{w_a,w_b\}$ in postorder, that is, $w_b$ first and
$w_a$ second. We have $t(w_b)=0$, so we put $\omega(w_b)=\max\{\omega(w_c),\omega(b)\}=1$, and
$\omegaExclHyb(w_b)=\max\{\omegaExclHyb(w_c),\omega(b)\}=1$ (Line~\ref{l:if-tw-0}). Since
$t(w_a)=1$, we put $\omega(w_a)=\omega(w_b)+\omega(a)=2$, and
$\omegaExclHyb(w_a)=\omegaExclHyb(w_b)+\omega(a)=2$ (Line~\ref{l:if-tw-1}). Finally, we consider the
vertex $\rho_C=v_1$. We have $t(v_1)=0$, so we go to Line~\ref{l:if1}. We first define
$\omega(v_1)=\max\{\omega(w_a),\omega(w)\}=4$ (Line~\ref{l:parallel}). Since
$\argmax\{\omega(w),\omega(w_a)\}=w$ and $w \in P^1$, we mark $w$ as \texttt{active}
(Line~\ref{l:activate-picked}), and we call  the procedure \textsc{Activate}$\left((N,t), \{P^1\},
\omega, \omegaExclHyb, 0, 0, 0 \right)$ (Line~\ref{l:activate-0}). 

We are now in the procedure \textsc{Activate}. We have $label_{\rho_C}=0$ and thus, go
to the for-loop in Line \ref{l:active:0:P}. Here, we consider the elements of $V(P^1) \setminus
\{\rho_C\}=\{w,w_c\}$ in postorder, that is $w_c$ first and $w$ second. We have $t(w_c)=0$ and $w_c$
does not have a child in $C$, so we mark $c$ as \texttt{active} (Line~\ref{l:active:0:0}). Since
$t(w)=1$, we mark both $v_2$ and $w_c$ as \texttt{active} (Line~\ref{l:active:0:1}). After this, we
exit the procedure.

We are now back in Algorithm~\ref{alg:clique}. Since all vertices of $V(\MDT_G)\setminus
L(\MDT_G)$ have now been processed, we are in Line~\ref{l:omega} and ready to compute the set
$\Omega$. The vertices marked as \texttt{active} are $g,h,u,u_d,v_3,e,d,w,c,v_2$ and $w_c$. Therefore, the set $\Omega$ computed at Line~\ref{l:omega} is $\{g,h,d,c\}$.
Note that although $e$ is also marked as \texttt{active}, its parent $u_e$ is not, so $e$ is not
added to $\Omega$. The algorithm stops here and returns $G[\Omega]=G[\{g,h,d,c\}]$ and
$\omega(\rho_N)=\omega(v_1)=4$. One can verify that $G[\{g,h,d,c\}]$ is indeed a maximum clique of
$G$. In particular, $\omega(v_1)$ corresponds to the size of a maximum clique in $G$.
\end{example}

\begin{proposition}\label{prop:clique-algo1}
Algorithm \ref{alg:clique} correctly computes the clique number $\omega(G)$ of
\gatex graphs $G$. In particular, if $(N,t)$ is a pvr-network of $G$ used
in Algorithm \ref{alg:clique}, then $\omega(v) = \omega(G[L(N(v))])$ for all $v\in V(N)$. 
In addition, if $v$ is contained in a cycle $C$ of $N$ and $v\neq \rho_C$, then
$\omegaExclHyb(v)=\omega(G[L(N(v))\setminus L(N(\eta_C)])$. 
\end{proposition}
\begin{proof}
Let $G=(V,E)$ be the input \gatex graph for Algorithm \ref{alg:clique}. In order
to show that $\omega(G)$ is correctly computed, we can ignore all Lines in
Algorithm \ref{alg:clique} where vertices are marked as active and where the
procedure \textsc{Activate} is called. We start in Line \ref{l:MDT-pvr} with
computing $(\MDT_G,t_G)$ and a pvr-network $(N,t)$ of $G$. In what follows, let
$L_w \coloneqq L(N(w))$ for $w\in V(N)$. Furthermore, for a vertex $w\in
V(\MDT_G)$, let $M_w\coloneqq L(\MDT_G(w))$ denote the module of $G$
``associated'' with $w$. To recall, $V(\MDT_G)\subseteq V(N)$.

In Line \ref{l:omega-leaves}, we initialize $\omega(v) = 1$ for all leaves $v\in L(\MDT_G) = L(N) = V$ 
and, thus, correctly capture the size $\omega(G[L_v])=\omega(v)$ of
a maximum clique in $G[L_v]\simeq K_1$. We then continue to traverse the
remaining vertices in $\MDT_G$ in postorder. This ensures that whenever we reach
a vertex $v$ in $\MDT_G$, all its children have been processed. We show now that
$\omega(v)\coloneqq \omega(G[M_v])$ is correctly computed for all $v\in
V(\MDT_G)$. Let $v$ be the currently processed vertex in Line \ref{l:forV}. By
induction, we can assume that the children $u$ of $v$ in $\MDT_G$ satisfy
$\omega(u) = \omega(G[M_u])$. We consider now the cases for $t(v)\in
\{0,1,\mathrm{prime}\}$. 

\emph{Case $t_G(v)=0$:} In this case, $\omega(v)$ is defined as $\max_{w\in \child_{\MDT_G}(v)}
\{\omega(w)\}$ in Line~\ref{l:v-parallel}. Lemma~\ref{lm:mdtclique}, together with the fact that the
children of $v$ is $\MDT_G$ are precisely the children of $v$ in $N$ (Observation~\ref{obs:ancest}),
implies that $\omega(G[M_v])=\max_{w\in \child_{\MDT_G}(v)}\{\omega(G[M_w])\}$. Therefore,
$\omega(v)=\max_{w\in \child_{\MDT_G}(v)}\{\omega(w)\}=\max_{w\in
\child_{\MDT_G}(v)}\{\omega(G[M_w])\}=\omega(G[M_v])$ follows.

\emph{Case $t_G(v)=1$:} In this case, $\omega(v)$ is defined as $\sum_{w\in \child_{\MDT_G}(v)}
\omega(w)$ in Line~\ref{l:v-series}. Lemma~\ref{lm:mdtclique}, together with the fact that the
children of $v$ is $\MDT_G$ are precisely the children of $v$ in $N$ (Observation~\ref{obs:ancest}),
implies that we have $\omega(G[M_v])=\sum_{w\in \child_{\MDT_G}(v)}\{\omega(G[M_w])\}$. Therefore,
$\omega(v)=\sum_{w\in \child_{\MDT_G}(v)}\omega(w)=\sum_{w\in
\child_{\MDT_G}(v)}\omega(G[M_w])=\omega(G[M_v])$ follows.

\emph{Case $t_G(v)=\mathrm{prime}$:} In this case, $M\coloneqq M_v$ is a prime
module of $G$ and $v$ is locally replaced by a cycle $C\coloneqq C_{M}$ with
root $\rho_C=v$ according to Def.\ \ref{def:pvr} and we have $M =
L(\MDT_G(v))=L_{\rho_C}$ (cf.\ Obs.\ \ref{obs:properties_C_pvr}). Although
$\rho_C=v$, we will distinguish between them to better keep track as whether we are
working in $\MDT_G$ or $N$. Let $P^1$ and $P^2$ be the two sides of $C$. 
By Obs.\ \ref{obs:properties_C_pvr}, all vertices
$w\neq \rho_C$ in $C$ have exactly one child $u''$ that is not in $C$. By
construction of $(N,t)$, each of those childs $u''$ is a child of $v$ in
$\MDT_G$. By induction assumption, we can assume that $\omega(u'')$ correctly
captures the size of $\omega([G[M_{u''}])=\omega([G[L_{u''}])$. Out task is now
to determine the clique number $\omega(v) \coloneqq \omega(G[M])$ of $G[M]$. In
the following, we will record two values $\omega(w)$ and $\omegaExclHyb(w)$ for
the vertices $w\neq \rho_C$ in $C$ to capture the size $\omega(w)$ of a maximum
clique in $G[L_w]$ and the size $\omegaExclHyb(w)$ of a maximum clique in
$G[L_w\setminus L_\eta]$.

We start in Line \ref{l:eta} with the the unique hybrid-vertex $\eta=\eta_C$ of
$C$. By Obs.\ \ref{obs:properties_C_pvr}, $\eta$ has precisely one child $u$
and, therefore, $L_\eta=L_u$. Hence, $\omega(\eta)\coloneqq \omega(u) =
\omega(G[L_u])$ and, since $G[L_u]=G[L_\eta]$, $\omega(\eta) =
\omega(G[L_\eta])$ is correctly determined in Line \ref{l:omega-eta}. Moreover,
$\omegaExclHyb(\eta) \coloneqq 0$ is correctly determined as there is no clique
in $G[L_\eta \setminus L_\eta]$. 

In Line \ref{l:init-innerC-start} - \ref{l:init-innerC-end}, we consider all
vertices $w\in V(C)\setminus\{\rho_C,\eta\}$ in a bottom-up order. By Obs.\
\ref{obs:properties_C_pvr}, $w$ has precisely two children $u'$ and $u''$ where
$u'$ is located on $C$ while $u''$ is not and it holds that $L_{u'}\cap
L_{u''}=\emptyset$. By the post-ordering, we start with one of the parents
of $\eta$ located in $C$.

Let $w$ be a parent of $\eta$ that is located in $P^i$ for some $i\in \{1,2\}$ for which
$u'=\eta$. Since $w$ is a parent of $\eta$ in $C$ it holds that $L_w=L_{\eta} \cup L_{u''}$
and $L_w \setminus L_{\eta}=L_{u''}$ and thus, in particular, $\omega(G[L_w \setminus
L_{\eta}])=\omega(G[L_{u''}])$. Assume that $t(w)=0$ (Line \ref{l:if-tw-0}). In this case, we put
$\omega(w) = \max\{\omega(\eta),\omega(u'')\}$ and $\omegaExclHyb(w)=\max\{\omegaExclHyb(eta),
\omega(u'')\} = \max\{0, \omega(u'')\} = \omega(u'')$. By our induction hypothesis,
$\omega(u'')=\omega(G[L_{u''}])$, so since $\omega(G[L_w \setminus L_{\eta}])=\omega(G[L_{u''}])$,
$\omegaExclHyb(w)=\omega(G[L_w \setminus L_{\eta}])$ follows. Moreover, since $t(w)=0$,
Lemma~\ref{lm:mdtclique} implies that $\omega(G[L_w])=\max
\{\omega(G[L_\eta]),\omega(G[L_{u''}])\}$. By our induction hypothesis,
$\omega(\eta)=\omega(G[L_\eta])$ and $\omega(u'')=\omega(G[L_{u''}])$, so $\omega(w) =\omega(G[w])$
follows. Assume now that $t(w)=1$ (Line \ref{l:if-tw-1}). In this case, we have we put
$\omega(w)=\omega(u')+\omega(u'')$ and $\omegaExclHyb(w)=\omegaExclHyb(u') + \omega(u'') = 0 +
\omega(u'')$. As in the previous case, $\omega(u'')=\omega(G[L_{u''}])$, together with $\omega(G[L_w
\setminus L_{\eta}])=\omega(G[L_{u''}])$, implies that $\omegaExclHyb(w)=\omega(G[L_w \setminus
L_{\eta}])$. Moreover, since $t(w)=1$, Lemma~\ref{lm:mdtclique} implies that
$\omega(G[L_w])=\omega(G[L_\eta])+\omega(G[L_{u''}])$. By our induction hypothesis,
$\omega(\eta)=\omega(G[L_\eta])$ and $\omega(u'')=\omega(G[L_{u''}])$, so $\omega(w) =\omega(G[w])$
follows.

Suppose now that $w\in V(C)\setminus(\{\rho_C,\eta\}\cup \parent(\eta))$ is the currently processed
vertex. Note that both children $u'$ and $u''$ of $w$ have already been processed
and we can assume by the latter arguments and by induction that $\omega(u') =
\omega(G[L_{u'}])$, $\omegaExclHyb(u') = \omega(G[L_{u'}\setminus L_\eta])$, and
$\omega(u'') = \omega(G[L_{u''}])$.

Assume that $t(w)=0$ (Line \ref{l:if-tw-0}). Then, we put $\omega(w) =
\max\{\omega(u'),\omega(u'')\}$ and by similar argument as used in the previous case, $\omega(w) =
\omega(G[L_w])$ is correctly computed. Consider now $\omegaExclHyb(w)=\max\{\omegaExclHyb(u'),
\omega(u'')\}$. By Obs.\ \ref{obs:properties_C_pvr} it holds that $\lca_N(x,y) = w$ for all $x\in
L_{u'}\setminus L_\eta\subseteq L_{u'}$ and $y\in L_{u''}$. This and $t(w)=0$ implies that there are
no edges between vertices in $G[L_{u'}\setminus L_\eta] $ and $G[L_{u''}]$. Hence, $G[L_w\setminus
L_\eta] = G[L_{u'}\setminus L_\eta]\cupdot G[L_{u''}]$. This together with $(L_w\setminus
L_\eta)\cap L_{u''}=\emptyset$ implies that $\omega(G[L_w\setminus L_\eta]) =
\max\{\omega(G[L_{u'}\setminus L_\eta]), \omega(G[L_{u''}])\}$. By our induction hypothesis,
$\omegaExclHyb(u')=\omega(G[L_{u'}\setminus L_\eta])$ and $\omega(u'')=\omega(G[L_{u''}])$, so
$\omegaExclHyb(w) =\omega(G[L_w \setminus L_{\eta}])$ follows. Suppose now that $t(w)=1$ (Line
\ref{l:if-tw-1}). Then, we put $\omega(w) = \omega(u') + \omega(u'')$ and by similar argument as
used in the previous case ($w$ as a parent of $\eta$), $\omega(w) = \omega(G[L_w])$ is correctly
computed. Consider now $\omegaExclHyb(w)=\omegaExclHyb(u')+\omega(u'')$. Since $\lca_N(x,y) = w$ for
all $x\in L_{u'}\setminus L_\eta$ and $y\in L_{u''}$, and $t(w)=1$, all vertices in
$G[L_{u'}\setminus L_\eta] $ are adjacent to all vertices in $G[L_{u''}]$. Hence, $G[L_w\setminus
L_\eta] = G[L_{u'}\setminus L_\eta]\join G[L_{u''}]$ and therefore, $\omega(G[L_w\setminus L_\eta])
= \omega(G[L_{u'}\setminus L_\eta])+\omega(G[L_{u''}])$. By our induction hypothesis,
$\omegaExclHyb(u')=\omega(G[L_{u'}\setminus L_\eta])$ and $\omega(u'')=\omega(G[L_{u''}])$, so
$\omegaExclHyb(w) =\omega(G[L_w \setminus L_{\eta}])$ follows.

In summary, in Line \ref{l:init-innerC-start} - \ref{l:init-innerC-end} the
values $\omega(w)=\omega(G[L_w])$ and $\omegaExclHyb(w)=\omega(G[L_w\setminus
L_\eta])$ have been correctly computed for all $w\in V(C)\setminus\{\rho_C\}$.

In Line \ref{l:rho1} - \ref{l:rho2} we finally determine the value
$\omega(v)$. To recall, $v=\rho_C$ is the vertex in $(\MDT_G,t_G)$ with label
$t_G(v)=\mathrm{prime}$ and $M = L(\MDT_G(v))$ is a prime module in
$G$ for which $M = L(\MDT_G(v)) = L_{\rho_C}$ holds. Let $u'$ and $u''$ be the two children
of $\rho_C$ (cf.\ Obs.\ \ref{obs:properties_C_pvr}). It is an easy task to
verify that $G[L_{u'}] = G_i(M)$ and $G[L_{u''}] = G_j(M)$ with
$\{i,j\}=\{1,2\}$. W.l.o.g.\ assume that $i=1$ and $j=2$. By induction, 
we can assume that $\omega(u')=\omega(G[L_{u'}])$ and $\omega(u'')=\omega(G[L_{u''}])$
and, therefore, $\omega(u')= \omega(G_1(M))$ and $\omega(u'')= \omega(G_2(M))$.
Assume now that $t(\rho_C)=0$. In this case, we put in Line \ref{l:parallel},
$\omega(v) \coloneqq \max\{\omega(u'), \omega(u'')\}$. By the latter
arguments, $\omega(v)=\max\{\omega(G_1(M)), \omega(G_2(M))\}$. By Prop.\
\ref{prop:clique-Number-M-parallel}, $\omega(G[M]) = \max\{\omega(G_1(M)),
\omega(G_2(M))\}$. Hence, $\omega(v) =  \omega(G[M])$ has been correctly determined
Assume now that $t(\rho_C) = 1$. Put $\alpha\coloneqq
\omega(u')+\omega(u'')-\omega(\eta)$ and $\beta\coloneqq \omegaExclHyb(u') +
\omegaExclHyb(u'')$. By the latter arguments and induction assumption, $\alpha =
\omega(G_1(M)) + \omega(G_2(M)-\omega(G[L_\eta])$ and $\beta=
\omega(G[L_{u'}\setminus L_\eta]) + \omega(G[L_{u'}\setminus L_\eta]) =
\omega(G_1(M)-G[L_\eta]) + \omega(G_2(M)-G[L_\eta])$. This together with Prop.\
\ref{prop:clique-Number-M-series} implies that 
$\omega(v)=\max\{\alpha,\beta\} = \omega(G[M]$ has
been correctly determined in Line \ref{l:series-2}.
 
Hence, by induction, $\omega(\rho_{\MDT_G})$ captures the  
 size of a maximum clique in $G[M_{\rho_{\MDT_G}}]$. Since $M_{\rho_{\MDT_G}}= V$, 
 we have $\omega(\rho_{\MDT_G}) = \omega(G[V])=\omega(G)$, which completes the proof.
Even more, the arguments above imply that $\omega(v) = \omega(G[L_v])$ holds for all $v\in V(N)$
and, if $v$ is contained in a cycle $C$ of $N$ and $v\neq \rho_C$, then
$\omegaExclHyb(v)=G[L_v\setminus L_{\eta}]$. 
\end{proof}

\begin{proposition}\label{prop:clique-algo2}
Algorithm \ref{alg:clique} correctly computes a maximum clique in \gatex graphs.
\end{proposition}
\begin{proof} 
	Let $G=(V,E)$ be the input \gatex graph for Algorithm \ref{alg:clique} and $(N,t)$ be the
	pvr-network that explains $G$ and that is used in Algorithm \ref{alg:clique}. In what follows,
	put $L_w \coloneqq L(N(w))$ for $w\in V(N)$. Furthermore, for a vertex $w\in V(\MDT_G)$, let
	$M_w\coloneqq L(\MDT_G(w))$ denote the module of $G$ associated with $w$. By Prop.\
	\ref{prop:clique-algo1}, $\omega(v) = \omega(G[L_v])$ for all $v\in V(N)$ and, if $v$ is
	contained in a cycle $C$ of $N$ and $v\neq \rho_C$, then $\omegaExclHyb(v)=G[L_v\setminus
	L_{\eta_C}]$.

	In the following, we call a directed path in $N$ from $w$ to some leaf in $L_w$ an
	\texttt{active} \emph{$w$-path} if all vertices in $P$ distinct from $w$ are marked as
	\texttt{active}. Moreover, we say that \emph{Property ($\star$) is satisfied for a vertex $w \in V(N)$}
	if a maximum clique in $G[L_w]$ is induced by all those leaves in $L_w$ that can be reached from
	\texttt{active} $w$-paths. We show that all vertices in $V(\MDT_G)\subseteq V(N)$ satisfy \emph{Property ($\star$)}.
	Note that \emph{Property ($\star$)} is trivially satisfied for all leaves in $L(N)$. Let $v$ be the currently
	processed vertex in Line \ref{l:forV}. By induction, we can assume that the children $u$ of $v$
	in $\MDT_G$ satisfy \emph{Property ($\star$)}. We consider now the cases for $t_G(v)\in \{0,1,\mathrm{prime}\}$. 

	\emph{Case $t_G(v)=0$:} In this case, 
	it follows from Observation~\ref{obs:ancest} that the children of $v$ in $N$ are precisely the children of $v$ in $\MDT_G$, that is, $\child_N(v)=\child_{\MDT_G}(v)$. 
	By Lemma~\ref{lm:mdtclique}, every maximum clique in $G[L_v]$ must be located entirely in
	one of the subgraphs $G[L_w]$,
	$w\in \child_N(v)$, of $G[L_v]$. In this case, one of the children $w\in \child_N(v)$
	satisfying $\omega(w) = max \{\omega(z) \mid z\in \child_N(v)\}$ is marked as
	\texttt{active} (Line \ref{l:active-0}). By induction assumption, \emph{Property ($\star$)} holds for $w$ and,
	in particular, $w$ is now \texttt{active}. This and the fact that a maximum clique in $G[L_v]$ is
	located entirely in $G[L_w]$
	implies that \emph{Property ($\star$)} holds for $v$. 

	\emph{Case $t_G(v)=1$:} In this case, 
	it again follows from Observation~\ref{obs:ancest} that the children of $v$ in $N$ are precisely
	the children of $v$ in $\MDT_G$, that is, $\child_N(v)=\child_{\MDT_G}(v)$. By Lemma~\ref{lm:mdtclique}, a maximum clique in $G[L_v]$ is the join union of the maximum cliques in $G[L_w]$, $w\in \child_N(v)$. In this case, all
	children $w\in \child_N(v)$ are marked as \texttt{active} (Line \ref{l:active-1}). By induction
	assumption, \emph{Property ($\star$)} holds for all $w\in \child_N(v)$ and, in particular, all
	$w\in \child_N(v)$ are now \texttt{active}. Taken the latter arguments together,
	\emph{Property ($\star$)} holds for $v$.

	\emph{Case $t_G(v)=\mathrm{prime}$:} In this case, $M\coloneqq M_v$ is a prime module of $G$ and
	$v$ is locally replaced by a cycle $C\coloneqq C_{M}$ with unique hybrid $\eta\coloneqq \eta_C$
	and root $\rho_C=v$ according to Def.\ \ref{def:pvr}. Let $P^1$ and $P^2$ be the two sides of $C$
	and $u'\in P^1$ and $u''\in P^2$ be the two children of $\rho_C$ in $N$. By Prop.\
	\ref{prop:clique-algo1}, $\omega(u') = \omega(G[L_{u'}])$ and $\omega(u'') = \omega(G[L_{u''}])$.
	In $(N,t)$, we either have $t(\rho_C)=1$ or $t(\rho_C)=0$. 

	Assume first that $t(\rho_C)=0$. In Line \ref{l:parallel}, we put $\omega(v) \coloneqq
	\max\{\omega(u'), \omega(u'')\} = \max\{\omega(G[L_{u'}],\omega(G[L_{u''}])\}$. By Prop.\
	\ref{prop:clique-algo1}, $\omega(v) = \omega(G[L_{v}])$. We then pick in Line \ref{l:pick-w} one
	of the vertices $w=u'$ or $w=u''$ for which $\omega(w) = \max\{\omega(u'), \omega(u'')\}$ is
	satisfied and determine in Line \ref{l:pick-P} the side $P\in \{P^1,P^2\}$ of $C$ that contains
	$w$. Afterwards, the procedure \textsc{Activate}$\left((N,t), \{P\}, \omega, \omegaExclHyb, 0, 0,
	0\right)$ is called in Line \ref{l:activate-0}. We are now in the procedure
	\textsc{Activate}$\left((N,t), \mathcal{P}, \omega, \omegaExclHyb, \alpha, \beta,
	\textit{label}\_\rho_C \right)$. In this case, we have $\textit{label}\_\rho_C=0$ and are
	therefore in the \emph{for}-loop in Line \ref{l:active:0:P} of this procedure. Here, $\mathcal{P}
	= \{P\}$ and we traverse the vertices $w'$ in $P$ in postorder. By construction, each $w'$
	has exactly two children, one of them is located in $C$ and denoted by $u'$ while the other one
	is the one outside of $C$ and is denoted by $u''$ (Line \ref{l:active:0:u'u''}). By
	Observation~\ref{obs:ancest}, the child $u''$ of $w'$ is always a child of $v$ in $\MDT_G$. By our induction
	hypothesis, $u''$ satisfies \emph{Property ($\star$)}. We now show that $w'$ satisfies
	\emph{Property ($\star$)}. Note that the first vertex considered in the procedure
	\textsc{Activate} is $w'=\eta$. By Obs.\ \ref{obs:properties_C_pvr}, $\eta$ has precisely one
	child. One easily verifies that in both cases, $t(\eta)=0$ or $t(\eta)=1$, the unique child
	$u''$ of $\eta$ is marked as \texttt{active}. Since $u''$ satisfies \emph{Property ($\star$)},
	and $L_{\eta}=L_{u''}$, $\eta$ satifies \emph{Property ($\star$)}. Suppose now that $w'$ is
	distinct from $\eta$. By induction, we can assume that the child $u'$ of $w'$ in
	$C$ satisfies \emph{Property ($\star$)}. Note that we can use this assumption, since $u'$ is
	processed before $w'$ in the procedure \textsc{Activate} and since $\eta$ has been processed
	already. By Lemma~\ref{lm:mdtclique},
	$t(w')=0$ implies that every maximum clique in $G[L_{w'}]$ must be located entirely in one of
	$G[L_{u'}]$ or $G[L_{u''}]$. In the procedure \textsc{Activate} (Line \ref{l:active:0:0}), we
	mark the child $u$ of $w$ satisfying $\omega(u)=\max\{\omega(u'),\omega(u'')\}$ as
	\texttt{active}. By our induction hypothesis, property ($\star$) holds for $u$. This and the fact
	that $u$ is \texttt{active} and that there exists a maximum clique in $G[L_{w'}]$ located
	entirely in $G[L_u]$ implies that \emph{Property ($\star$)} holds for $w'$. If $t(w')=1$ then,
	by Lemma~\ref{lm:mdtclique}, a maximum clique in $G[L_{w'}]$
	is the join union of a maximum cliques in $G[L_{u'}]$ and a maximal clique in $G[L_{u''}]$. In
	this case, both $u'$ and $u''$ are marked as \texttt{active} in the procedure \textsc{Activate}
	(Line \ref{l:active:0:1}). By induction, \emph{Property ($\star$)} holds for both $u'$ and $u''$,
	and both of them are now \texttt{active}. The latter two arguments imply that  \emph{Property
	($\star$)} holds for $w'$. In particular, \emph{Property ($\star$)} holds for the chosen
	child $w$ of $v$ in $N$. Note that $w$ was marked as \texttt{active}
	in Alg.~\ref{alg:clique} (Line \ref{l:activate-picked}), while the other child of $v$ is
	not. Since by choice of $w$, $G[L_v]$ admits a maximum clique entirely contained in $G[L_w]$, it
	follows that $v$ satisfies \emph{Property ($\star$)}.

	Assume now that $t(\rho_C)=1$. In this case, we call
	\textsc{Activate}$\left((N,t), \{P^1,P^2\}, \omega, \omegaExclHyb, \alpha, \beta, 1\right)$
	in Alg.\ \ref{alg:clique} (Line \ref{l:activate-1})
	where  $\alpha = \omega(u')+\omega(u'')-\omega(\eta)$ and $\beta = \omegaExclHyb(u') + \omegaExclHyb(u'')$. 
	To recall, $u'\in P^1$ and $u''\in P^2$ are the two children of $\rho_C$ in $N$.
	As argued in the proof of Prop.\ \ref{prop:clique-algo1},  $\alpha =
	\omega(G_1(M)) + \omega(G_2(M)-\omega(G[L_\eta])$ and $\beta=\omega(G_1(M)-G[L_\eta]) + \omega(G_2(M)-G[L_\eta])$. 
	Since $t(\rho_C)=1$, we continue in Line \ref{l:active:1} of procedure \textsc{Activate}. 
	There are two cases, either $\alpha\leq \beta$ or $\alpha>\beta$. 

	Assume first that $\alpha\leq \beta$. In this case,
	Proposition~\ref{prop:clique-Number-M-series} implies that a maximum clique in $G[L_v]$ can be
	obtained by taking the join union of one maximum clique in $G[L_{v_1} \setminus L_{\eta}]$ and
	one one maximum clique in $G[L_{v_2}\setminus L_{\eta}]$, where $v_1$ and $v_2$ are the two
	children of $\rho_C = v$ in $N$. Hence, a a maximum clique in $G[L_v]$ is, in particular, a
	maximum clique in $G[L_v\setminus L_{\eta}]$ Since $\alpha\leq \beta$, we are in Line
	\ref{l:active:1:aLEQb} of the procedure \textsc{Activate}. For all vertices $w'$ in $P_1$ and
	$P_2$ distinct from $\rho_C$ and $\eta$, Observation~\ref{obs:ancest} implies that the unique
	child $u''$ of $w'$ outside of $C$ is a child of $v$ in $\MDT_G$. In particular, by our induction
	hypothesis, $u''$ satisfies \emph{Property ($\star$)}. We now proceed with showing that every
	$w'$ in $P_1$ and $P_2$ distinct from $\rho_C$ and $\eta$, satisfies the following amended
	version of \emph{Property ($\star$)}. Namely, we say that $w'$ satisfies \emph{Property
	($\star\star$)} if a maximum clique in $G[L_{w'} \setminus L_{\eta}]$ is induced by all leaves in
	$L_{w'}$ that can be reached from the \texttt{active} $w'$-paths. If $w'$ is a parent of $\eta$
	in $N$, then $L_{w'} \setminus L_{\eta}=L_{u''}$, and $u''$ is marked as \texttt{active} in .
	Since $u''$ satisfies \emph{Property ($\star$)} by our induction hypothesis, it follows that $w'$
	satisfies Property ($\star\star$) in Line \ref{l:active:1:parent-eta}. All remaining vertices
	$w' $in $P_1$ and $P_2$. i.e., those that are distinct from $\rho_C,\eta$ and its two unique
	parents $w_1$ and $w_2$ are now traversed in postorder (Line \ref{l:active:1:P}). Suppose now
	that $w'$ is one of these vertices and let $u'$ be the child
	of $w'$ in $C$. Since the vertices of $P_1$ and $P_2$ are processed in postorder in the procedure
	\textsc{Activate}, we may assume that $u'$ satisfies \emph{Property ($\star\star$)}. The
	latter is justified since the parents of $\eta$ have been processed and satisfy \emph{Property
	($\star\star$)}. If $t(w')=0$, then similar arguments as in the proof of
	Lemma~\ref{lm:mdtclique} imply that a maximum clique in $G[L_{w'} \setminus L_{\eta}]$ is
	contained either in $G[L_{u''}]$ or in $G[L_{u'} \setminus L_{\eta}]$. In the procedure
	\textsc{Activate} (Line~\ref{l:pick-u}), we mark $u'$ as active if $\omegaExclHyb(u') \geq
	\omega(u'')$, and we mark $u''$ as active otherwise. By induction, \emph{Property ($\star$)}
	holds for $u''$, and \emph{Property ($\star\star$)} holds for $u'$. This and the fact that there
	exists a maximum clique in $G[L_{u'}]$ located entirely in $G[L_{u'} \setminus L_{\eta}]$ (in
	case $\omegaExclHyb(u') \geq \omega(u'')$) or in $G[L_{u''}]$ (in case $\omega(u'') \geq
	\omegaExclHyb(u')$) implies that \emph{Property ($\star\star$)} holds for $w'$. If $t(w')=1$,
	then similar arguments as in the proof of Lemma~\ref{lm:mdtclique} imply that a maximum
	clique in $G[L_{w'}\setminus L_{\eta}]$ is the join union of a maximum clique in $G[L_{u'}
	\setminus L_{\eta}]$ and a maximum clique in $G[L_{u''}]$. In this case, both $u'$ and $u''$ are
	marked as \texttt{active} in the procedure \textsc{Activate} (Line \ref{l:active:1:active}). By
	our induction hypothesis, \emph{Property ($\star$)} holds for $u''$ and \emph{Property
	($\star\star$)} holds for $u'$. The latter two arguments imply that \emph{Property
	($\star\star$)} holds for $w'$. In particular, \emph{Property ($\star\star$)} holds for the
	children $v_1$ and $v_2$ of $v$ in $N$. Note that both $v_1$ and $v_2$ are marked as
	\texttt{active} (Line~\ref{l:active:1}). Moreover, $\eta$ is not marked as \texttt{active}, so
	for $i \in \{1,2\}$, all leaves that can be reached from $v_i$ via a path of \texttt{active}
	vertices are not descendants of $\eta$. It follows that for all leaves $x_1 \in L_{v_1}, x_2 \in
	L_{v_2}$ that can be reached by such a path, $\lca_N(x_1,x_2)=v$. Since $t(v)=1$ and $(N,t)$
	explains $G$, it follows that $\{x_1,x_2\}$ is an edge of $G$. Together with the fact that $v_1$
	and $v_2$ satisfy \emph{Property ($\star\star$)}, this implies that the set of leaves of $L_v$
	that can be reached from $v$ via a path of \texttt{active} vertices induces a clique of $G[L_v]$
	of size $\omega(G[L_{v_1} \setminus L_{\eta}])+\omega(G[L_{v_2} \setminus L_{\eta}])$. By
	Proposition~\ref{prop:clique-algo1}, $\omega(G[L_{v_1} \setminus L_{\eta}])+\omega(G[L_{v_2}
	\setminus L_{\eta}])=\omegaExclHyb(v_1)+\omegaExclHyb(v_2)=\beta$, and since $\beta \geq \alpha$,
	Proposition~\ref{prop:clique-Number-M-series} implies that the latter clique is a maximum clique
	in $L_v$. Therefore, $v$ satisfies \emph{Property ($\star$)}.

	Assume now that $\alpha > \beta$. In this case, Proposition~\ref{prop:clique-Number-M-series}
	implies that every maximum clique in $G[L_v]$ must contain vertices in $L_{\eta}$, i.e., we must
	subsequently build active parts while keeping \texttt{active} paths along $\eta$. Since $\alpha
	> \beta$, we are in Line \ref{l:active:1:a>b} of the procedure \textsc{Activate}, and we mark
	$\eta$ and its unique child as \texttt{active} (Line \ref{l:active:1:eta}) and proceed with
	traversing the vertices $w$ in $P_1$ and $P_2$ distinct from $\eta$ and $\rho_C$ in
	postorder (Line \ref{l:active:1:P2}). By Observation~\ref{obs:ancest}, for all such $w'$, the
	child $u''$ of $w'$ outside of $C$ is a child of $v$ in $\MDT_G$. In particular, by our induction
	hypothesis, $u''$ satisfies \emph{Property ($\star$)}. We now proceed to show that $w'$ satisfies
	\emph{Property ($\star$)}. Note that since the only child $u''$ if $\eta$ is \texttt{active}
	(Line~\ref{l:active:1:eta}), this is true for $w'=\eta$. Since the $w' \neq \eta$ vertices of
	$P_1$ and $P_2$ are processed in postorder, we may therefore assume that the child $u'$ of $w'$
	in $C$ satisfies \emph{Property ($\star$)}. By
	Lemma~\ref{lm:mdtclique}, $t(w')=0$ implies that every maximum clique in $G[L_{w'}]$ must be
	located entirely in one of $G[L_{u'}]$ or $G[L_{u''}]$. Moreover, by
	Proposition~\ref{prop:clique-Number-M-series}, every maximum clique in $G[L_v]$ contains vertices
	in $L_{\eta}$. As a consequence, since $L_{\eta} \subseteq L_{w'} \subseteq L_v$, a maximum
	clique in $G[L_{w'}]$ contains vertices in $L_{\eta}$. Since $L_{u''} \cap L_{\eta}=\emptyset$,
	the latter two arguments imply that every maximum clique in $G[L_{w'}]$ must be located entirely
	in $G[L_{u'}]$. Since $u'$ is marked as \texttt{active} (Line~\ref{l:active:1:w-active}), and
	$u'$ satisfies \emph{Property ($\star$)}, it follows that $w'$ satisfies \emph{Property
	($\star$)}. If $t(w')=1$ then, by Lemma~\ref{lm:mdtclique},
	a maximum clique in $G[L_{w'}]$ is the join union of a maximum clique in $G[L_{u'}]$ and a
	maximal clique in $G[L_{u''}]$. In this case, both $u'$ and $u''$ are marked as \texttt{active}
	in the procedure \textsc{Activate} (Lines~\ref{l:active:1:w-active} and \ref{l:active:1:w=1}). By
	our induction hypothesis, \emph{Property ($\star$)} holds for both $u'$ and $u''$, and both of
	them are now \texttt{active}. The latter two arguments imply that \emph{Property ($\star$)} holds
	for $w'$. In particular, \emph{Property ($\star$)} holds for the children $v_1$ and $v_2$ of $v$
	in $N$. Note that both $v_1$ and $v_2$ are marked as \texttt{active}
	(Line~\ref{l:active:1:w-active}). Note also that for $x \in L_{\eta}$, $x$ can be reached from
	$v_1$ via a path of \texttt{active} vertices if and only if $x$ can be reached from $v_1$ via a
	path of \texttt{active} vertices. Moreover, for all leaves $x_1 \in L_{v_1} \setminus L_{\eta},
	x_2 \in L_{v_2} \setminus L_{\eta}$, we have $\lca_N(x_1,x_2)=v$. Since $t(v)=1$ and $(N,t)$
	explains $G$, it follows that $\{x_1,x_2\}$ is an edge of $G$. Together with the fact that $v_1$
	and $v_2$ satisfy \emph{Property ($\star$)}, this implies that the set of leaves of $L_v$ that
	can be reached from $v$ via a path of \texttt{active} vertices induces a clique of $G[L_v]$ of
	size $\omega(G[L_{v_1}])+\omega(G[L_{v_2}])-\omega(G[L_{v_1} \cap
	L_{v_2}])=\omega(G[L_{v_1}])+\omega(G[L_{v_2}])-\omega(G[L_{\eta}])$. By
	Proposition~\ref{prop:clique-algo1},
	$\omega(G[L_{v_1}])+\omega(G[L_{v_2}])-\omega(G[L_{\eta}])=\omega(v_1)+\omega(v_2)-\omega(\eta)=\alpha$,
	and since $\alpha > \beta$, Proposition~\ref{prop:clique-Number-M-series} implies that the latter
	clique is a maximum clique in $L_v$. Therefore, $v$ satisfies \emph{Property ($\star$)}. 
\end{proof}

\begin{proposition}\label{prop:runtime-clique}
Algorithm \ref{alg:clique}  can be implemented to run in $O(|V|+|E|)$ time
with input $G=(V,E)$ 
\end{proposition}
\begin{proof} 
 We show now that Algorithm \ref{alg:clique} can be implemented to run in $O(|V|+|E|)$ time for a
given \gatex graph $G=(V,E)$. The modular decomposition tree $(\MDT_G,t_G)$ can be computed in
$O(|V|+|E|)$ time \cite{HP:10}. By \cite[Thm.\ 9.4 and Alg.\ 4]{HS:22}, the pvr-network $(N,t)$ of
$G$ can be computed within the same time complexity. Thus, Line \ref{l:MDT-pvr} takes $O(|V|+|E|)$
time. Initializing $\omega(v)\coloneqq 1$ for all leaves $v$ (and thus, the vertices of $G$) in Line
\ref{l:omega-leaves} can be done in $O(|V|)$ time.

 Note that $V$ is the leaf set of $\MDT_G$. We then traverse each of the $O(|V|)$ non-leaf
vertices in $(\MDT_G,t_G)$ in postorder starting in Line \ref{l:forV}. To simplify the arguments
and to establish the runtime, we put $W\coloneqq V(\MDT_G)\setminus V$ and partition the vertices in $W$ into $W_P\cupdot (W\setminus W_P)$ where $W_P$ contains
all vertices $v$ with $t(v)=\mathrm{prime}$. Moreover, we denote with $\deg_H(v)$ the number of
edges incident to $v$ in some DAG $H$.
 
 Note that for $v\in W\setminus W_P$ we have $\deg_N(v)=\deg_{\MDT_G}(v)$. All vertices $v\in
W\setminus W_P$ are processed in Line \ref{l:v-parallel} and \ref{l:active-0} as well as in Line
\ref{l:v-series} and \ref{l:active-1}. It is an easy task to verify that the respective two steps
take $O(\deg_N(v))=O(\deg_{\MDT_G}(v))$ time for each of the vertices in $W\setminus W_P$. Hence,
processing all vertices in $W\setminus W_P$ can be done in $O(\sum_{v\in W\setminus W_P}
\deg_{\MDT_G}(v)) = O(|E(\MDT_G)|) = O(|V(\MDT_G)| = O(|V|)$ time. 
 
 Now, consider the vertices in $W_P$ which are processed in Line \ref{l:v-prime}-\ref{l:rho2}. Note
first that the sides $P^1$ and $P^2$ of $C$ can be determined in $O(|V(C)|)$ time in Line
\ref{l:sides}. Moreover, it is easy to verify that, for each $v\in W_P$, all other individual steps
starting at Line \ref{l:v-prime} can be done in constant time each processed vertex has precisely two
children, except execution of the procedure \textsc{Activate} which takes $O(|V(C)|)$ time for each
individual call. For each $v\in W_P$, \textsc{Activate} is called once. Each
$v\in W_P$ is associated with the unique cycle $C^v\coloneqq C_M$ with $M=L(\MDT_G(v))$. Taken
together the latter arguments, for a given prime vertex $v$, Line \ref{l:v-prime} -
\ref{l:end-v-prime} have runtime $O(|V(C^v)|+|E(C^v)|) = |V(C^v)|$. Note that each cycle $C$ has, by
definition of pvr-networks, no vertex in common with every other cycles. Hence, processing all
vertices in $W_P$ can be done in $\sum_{v\in W_P} O(|V(C^v)|) = O(|V(N)|)$ 
 
 By \cite[Prop.\ 1]{CRV07}, we have $O(|V(N)|)=O(|V|)$. Hence, the overall time-complexity of
Algorithm \ref{alg:clique} is bounded by the time-complexity to compute $(\MDT_G,t_G)$ and $(N,t)$
in Line \ref{l:MDT-pvr} and is, therefore, $O(|V|+|E|)$ time.
\end{proof}

We consider now the problem of determining the independence number $\alpha(G)$ as well as a maximum
independent set of \gatex graphs $G$. Suppose that a \gatex graph $G$ is explained by the network
$(N,t)$ and let $\overline t\colon V(N)\to \{0,1,\dot\}$ where $\overline t(v)=\odot$ for all leaves
$v$ of $N$ and $\overline t(v)=1 $ if and only if $t(v)=0$. Since $L(N)=V(G)$ and by \cite[Prop.\
1]{CRV07}, we have $O(|V(N)|)=O(|V(G)|)$ and thus, this labeling can be computed in $O(|V(G)|)$
time. It is easy to verify that $(N,\overline t)$ explains the complement $\overline G$ of $G$. The
latter arguments imply that the complement of every \gatex graph is a \gatex graph as well. Since
maximum cliques in $\overline G$ are precisely the maximum independent sets in $G$, the latter
arguments together with Prop.~\ref{prop:clique-algo2} and \ref{prop:runtime-clique} imply

\begin{theorem}
A maximum clique and a maximum independent set can be computed in linear-time for \gatex graphs. 
\end{theorem}

\bibliographystyle{plain}
\bibliography{pc}

\begin{thebibliography}{10}

\bibitem{BSY:99}
Amir Ben-Dor, Ron Shamir, and Zohar Yakhini.
\newblock Clustering gene expression patterns.
\newblock {\em Journal of Computational Biology}, 6(3-4):281--297, 1999.
\newblock PMID: 10582567.

\bibitem{BD98}
Sebastian B{\"o}cker and Andreas W.~M. Dress.
\newblock Recovering symbolically dated, rooted trees from symbolic
  ultrametrics.
\newblock {\em Advances in Mathematics}, 138(1):105--125, 1998.

\bibitem{CRV07}
G.~Cardona, F.~Rossell\'o, and G.~Valiente.
\newblock Comparison of tree-child phylogenetic networks.
\newblock {\em {IEEE}/{ACM} {T}ransactions on {C}omputational {B}iology and
  {B}ioinformatics}, 6:552--569, 2007.

\bibitem{CHVATAL198463}
V.~Chvátal.
\newblock Perfectly ordered graphs.
\newblock In C.~Berge and V.~Chvátal, editors, {\em Topics on Perfect Graphs},
  volume~88 of {\em North-Holland Mathematics Studies}, pages 63--65.
  North-Holland, 1984.

\bibitem{Corneil:81}
D.~G. Corneil, H.~Lerchs, and L~K Stewart~Burlingham.
\newblock Complement reducible graphs.
\newblock {\em Discr. Appl. Math.}, 3:163--174, 1981.

\bibitem{Corneil:85}
D.~G. Corneil, Y.~Perl, and L.~K. Stewart.
\newblock A linear recognition algorithm for cographs.
\newblock {\em SIAM Journal on Computing}, 14(4):926--934, 1985.

\bibitem{DGC:01}
Elias Dahlhaus, Jens Gustedt, and Ross~M McConnell.
\newblock Efficient and practical algorithms for sequential modular
  decomposition.
\newblock {\em Journal of Algorithms}, 41(2):360 -- 387, 2001.

\bibitem{ER1:90}
A~Ehrenfeucht and G~Rozenberg.
\newblock Theory of 2-structures, part {I}: Clans, basic subclasses, and
  morphisms.
\newblock {\em Theor. Comp. Sci.}, 70:277--303, 1990.

\bibitem{ER2:90}
A~Ehrenfeucht and G~Rozenberg.
\newblock Theory of 2-structures, part {II}: Representation through labeled
  tree families.
\newblock {\em Theor. Comp. Sci.}, 70:305--342, 1990.

\bibitem{EHMS:94}
Andrzej Ehrenfeucht, Harold~N. Gabow, Ross~M. Mcconnell, and Stephen~J.
  Sullivan.
\newblock {An O($n^2$) Divide-and-Conquer Algorithm for the Prime Tree
  Decomposition of Two-Structures and Modular Decomposition of Graphs}.
\newblock {\em Journal of Algorithms}, 16(2):283--294, 1994.

\bibitem{garey1979computers}
Michael~R Garey and David~S Johnson.
\newblock {\em Computers and intractability}, volume 174.
\newblock freeman San Francisco, 1979.

\bibitem{HP:10}
M.~Habib and C.~Paul.
\newblock A survey of the algorithmic aspects of modular decomposition.
\newblock {\em Computer Science Review}, 4(1):41 -- 59, 2010.

\bibitem{Hellmuth:20b}
Marc Hellmuth, Adrian Fritz, Nicolas Wieseke, and Peter~F. Stadler.
\newblock Cograph editing: Merging modules is equivalent to editing {$P_4$'s}.
\newblock {\em Art Discr. Appl. Math.}, 3:\#P2.01, 2020.

\bibitem{HSS:22cluster}
Marc Hellmuth, David Schaller, and Peter~F. Stadler.
\newblock Clustering systems of phylogenetic networks.
\newblock {\em Theory in Biosciences}, 142(4):301--358, 2023.

\bibitem{HS:22}
Marc Hellmuth and Guillaume~E. Scholz.
\newblock From modular decomposition trees to level-1 networks:
  Pseudo-cographs, polar-cats and prime polar-cats.
\newblock {\em Discrete Applied Mathematics}, 321:179--219, 2022.

\bibitem{HS-GatexLinTime:23}
Marc Hellmuth and Guillaume~E. Scholz.
\newblock Linear time algorithms for {NP}-hard problems restricted to {GaTEx}
  graphs.
\newblock In Weili Wu and Guangmo Tong, editors, {\em Computing and
  Combinatorics}, pages 115--126. Springer Nature Switzerland, Cham, 2024.

\bibitem{HS-GatexForbSubg:23}
Marc Hellmuth and Guillaume~E. Scholz.
\newblock Resolving prime modules: The structure of pseudo-cographs and
  galled-tree explainable graphs.
\newblock {\em Discrete Applied Mathematics}, 343:25--43, 2024.

\bibitem{HSW:16}
Marc Hellmuth, Peter~F Stadler, and Nicolas Wieseke.
\newblock The mathematics of xenology: Di-cographs, symbolic ultrametrics,
  2-structures and tree-representable systems of binary relations.
\newblock {\em Journal of Mathematical Biology}, 75(1):199--237, 2017.

\bibitem{HS18}
K.~T. Huber and G.~E. Scholz.
\newblock Beyond representing orthology relations with trees.
\newblock {\em Algorithmica}, 80(1):73--103, 2018.

\bibitem{Luce1949}
R.~Duncan Luce and Albert~D. Perry.
\newblock A method of matrix analysis of group structure.
\newblock {\em Psychometrika}, 14(2):95--116, Jun 1949.

\bibitem{Marx2004GRAPHCP}
D{\'a}niel Marx.
\newblock Graph colouring problems and their applications in scheduling.
\newblock {\em Periodica Polytechnica Electrical Engineering}, 48:11--16, 2004.

\bibitem{CS:99}
Ross~M. McConnell and Jeremy~P. Spinrad.
\newblock Modular decomposition and transitive orientation.
\newblock {\em Discrete Mathematics}, 201(1-3):189 -- 241, 1999.

\bibitem{MIDDENDORF1990327}
Matthias Middendorf and Frank Pfeiffer.
\newblock On the complexity of recognizing perfectly orderable graphs.
\newblock {\em Discrete Mathematics}, 80(3):327--333, 1990.

\bibitem{SCHS:24}
Ameera~Vaheeda Shanavas, Manoj Changat, Marc Hellmuth, and Peter~F. Stadler.
\newblock Unique least common ancestors and clusters in directed acyclic
  graphs.
\newblock In Subrahmanyam Kalyanasundaram and Anil Maheshwari, editors, {\em
  Algorithms and Discrete Applied Mathematics}, pages 148--161, Cham, 2024.
  Springer Nature Switzerland.

\bibitem{VL:03}
Victor Spirin and Leonid~A. Mirny.
\newblock Protein complexes and functional modules in molecular networks.
\newblock {\em Proceedings of the National Academy of Sciences},
  100(21):12123--12128, 2003.

\bibitem{TCHP:08}
Marc Tedder, Derek Corneil, Michel Habib, and Christophe Paul.
\newblock Simpler linear-time modular decomposition via recursive factorizing
  permutations.
\newblock In {\em Automata, Languages and Programming}, volume 5125 of {\em
  Lecture Notes in Computer Science}, pages 634--645. Springer Berlin
  Heidelberg, 2008.

\bibitem{TurauWeyer:2015}
Volker Turau and Christoph Weyer.
\newblock {\em Algorithmische Graphentheorie}.
\newblock De Gruyter, Berlin, München, Boston, 2015.

\end{thebibliography}

\end{document}